\documentclass[%showpacs,
showkeys,12pt,
preprint,preprintnumbers,nofootinbib,
groupedaddress,superscriptaddress,amsmath,amssymb]{revtex4}
\usepackage{graphicx}% Include figure files
\usepackage{dcolumn}% Align table columns on decimal point
\usepackage{bm}% bold math
\usepackage{amssymb}
\usepackage{amsmath}
\usepackage{epsfig}    
\usepackage{color}
\usepackage{slashed}
\usepackage{hhline}
\def\be{\begin{equation}}
\def\ee{\end{equation}}
\newcommand{\bea}{\begin{eqnarray}}
\newcommand{\eea}{\end{eqnarray}}
\newcommand{\nn}{\nonumber}

\numberwithin{equation}{section}

\begin{document}

%%%%%%%%%
\title{Two-loop Induced Majorana Neutrino Mass\\
in
 a Radiative{ly} Induced  Quark and Lepton  
  Mass Model }
\preprint{KIAS-P16068}
 \author{Takaaki Nomura}
\email{nomura@kias.re.kr}
\affiliation{School of Physics, KIAS, Seoul 130-722, Korea}

\author{Hiroshi Okada}
\email{macokada3hiroshi@cts.nthu.edu.tw}
\affiliation{Physics Division, National Center for Theoretical Sciences, Hsinchu, Taiwan 300}

\date{\today}

\begin{abstract}
A two-loop induced radiative neutrino model is proposed as an extension of our previous work in which the first and second generation standard model fermion masses are generated at one-loop level in both quark and lepton sectors. Then we discuss current neutrino oscillation data, lepton flavor violations, muon anomalous magnetic moment, and a bosonic dark matter candidate, for both the normal and inverted neutrino mass hierarchy.
Our numerical analysis shows that less hierarchical Yukawa coupling constants can fit the experimental data with TeV scale dark matter.
\end{abstract}
\maketitle
\newpage

\section{Introduction}
In the standard model (SM), neutrinos and  dark matter (DM) candidates are not involved in if the neutrinos could be Majorana property within the renormalizable theory.
Radiatively induced mass models are one of the attractive candidates to accommodate these fields, and a lot of groups are trying to establish such models. For example, one-loop induced neutrino mass models are found in Refs.~\cite{a-zee, Cheng-Li, Pilaftsis:1991ug, Ma:2006km}, subsequently several variations has been achieved as found in~\cite{Gu:2007ug, Sahu:2008aw, Gu:2008zf, AristizabalSierra:2006ri, Bouchand:2012dx, McDonald:2013hsa, Ma:2014cfa, Kajiyama:2013sza, Kanemura:2011vm, Kanemura:2011jj, Kanemura:2011mw, Schmidt:2012yg, Kanemura:2012rj, Farzan:2012sa, Kumericki:2012bf, Kumericki:2012bh, Ma:2012if, Gil:2012ya, Okada:2012np, Hehn:2012kz, Dev:2012sg, Kajiyama:2012xg, Toma:2013zsa, Kanemura:2013qva, Law:2013saa, Baek:2014qwa, Kanemura:2014rpa, Fraser:2014yha, Vicente:2014wga, Baek:2015mna, Merle:2015gea, Restrepo:2015ura, Merle:2015ica, Wang:2015saa, Ahn:2012cg, Ma:2012ez, Hernandez:2013dta, Ma:2014eka, Ma:2014yka, Ma:2015pma, Ma:2013mga, radlepton1, Okada:2014nsa, Brdar:2013iea, Okada:2015kkj, Bonnet:2012kz, Joaquim:2014gba, Davoudiasl:2014pya, Lindner:2014oea, Okada:2014nea, Mambrini:2015sia, Boucenna:2014zba, Ahriche:2016acx, Fraser:2015mhb, Fraser:2015zed, Adhikari:2015woo, Okada:2015vwh, Ibarra:2016dlb, Arbelaez:2016mhg, Ahriche:2016rgf, Lu:2016ucn, Kownacki:2016hpm, Ahriche:2016cio, Ahriche:2016ixu, Ma:2016nnn, Nomura:2016jnl, Hagedorn:2016dze, Antipin:2016awv, Nomura:2016emz, Gu:2016ghu, Guo:2016dzl, Hernandez:2015hrt, Megrelidze:2016fcs, Cheung:2016fjo, Seto:2016pks, Lu:2016dbc}.
%%%
Two-loop, three-loop, and four-loop models are respectively found in Refs.~\cite{2-lp-zB, Babu:2002uu, AristizabalSierra:2006gb, Nebot:2007bc, Schmidt:2014zoa, Herrero-Garcia:2014hfa, Long:2014fja, VanVien:2014apa, Aoki:2010ib, Lindner:2011it, Baek:2012ub, Aoki:2013gzs, Kajiyama:2013zla, Kajiyama:2013rla, Baek:2013fsa, Okada:2014vla, Okada:2014qsa, Okada:2015nga, Geng:2015sza, Kashiwase:2015pra, Aoki:2014cja, Baek:2014awa, Okada:2015nca, Sierra:2014rxa, Nomura:2016rjf, Nomura:2016run, Bonilla:2016diq, Kohda:2012sr, Dasgupta:2013cwa, Nomura:2016ask, Nomura:2016dnf, Liu:2016mpf}, ~\cite{Krauss:2002px, Aoki:2008av, Gustafsson:2012vj, Ahriche:2014xra, Ahriche:2014cda, Ahriche:2014oda, Okada:2014oda, Hatanaka:2014tba, Jin:2015cla, Culjak:2015qja, Okada:2015bxa, Geng:2015coa, Ahriche:2015wha, Nishiwaki:2015iqa, Okada:2015hia, Ahriche:2015loa, Kajiyama:2013lja, King:2014uha, Kanemura:2015bli, Okada:2016rav, Ko:2016sxg, Nomura:2016vxr, Thuc:2016qva, Cherigui:2016tbm, Nomura:2016ezz, Cheung:2016ypw, Cheung:2016frv, Gu:2016xno}, and \cite{Nomura:2016fzs, Nomura:2016seu}.

%{\color{red} 
Previously we proposed a model in which first and second generation quark and lepton masses are radiatively generated at one-loop level for understanding the mass hierarchy in quark and charged lepton sector~\cite{Nomura:2016emz}. 
In this model, Dirac type neutrino mass terms are generated in the same way as the other sectors which require a tiny Yukawa coupling as $O(10^{-13} - 10^{-12})$ due to smallness of neutrino masses.
In this paper, we extend the model where the Majorana type active neutrinos are realized at two-loop level. %}
%based on our previous work, in which neutrinos were Dirac type and all the DM fermions are generated at one-loop level.
Thus the more natural hierarchy between neutrinos and the other SM fermion sectors can be achieved. 
Moreover, when neutrinos are Majorana type, it might be clarified through the experiments searching for neutrinoless double beta decay. 
 Also one can predict one of the three neutrinos is massless fermion, because one of the Yukawa couplings is anti-symmetric matrix. 
 This property is the same as the preceding work, i.e., Zee model found in the first reference of the one-loop model.  Due to the property, different patterns of allowed regions 
are obtained depending on normal hierarchy (NH) or inverted hierarchy (NH) of the neutrino masses.
In addition, several bosonic DM candidates are involved in and can be detected by direct or indirect detection searches.

This paper is organized as follows.
In Sec.~II, we show our model,  including neutrino sector, LFVs, muon anomalous magnetic moment, and bosonic DM candidate to explain direct detection and relic density.
In Sec.~III, we have a numerical analysis, and show some results.
We conclude and discuss in Sec.~IV.
%In appendices, we show the explicit Higgs potential and ...

%\newpage

%%%%%%%%%%%%%%%%%%%%%%%%%%%%%%%%%%%%%

 \begin{widetext}
\begin{center} 
\begin{table}%[tbc]
%\begin{tiny}
\begin{tabular}{|c||c|c|c|c|c|c||c|c|c|c|c|c|c|c|}\hline\hline  
&\multicolumn{6}{c||}{Leptons} & \multicolumn{2}{c|}{Bosons(VEVs$\neq$0)} & \multicolumn{4}{c|}{Bosons(VEVs$=$0)} \\\hline
Fields
& ~$L_L^\alpha$~ & ~$e_R^i$ ~ & ~$\tau_R$ ~& ~$L'^i_{L,R}$~& ~$N_R^i$~& ~$N_L^i$~ & ~$\Phi$~ &~ $\varphi$ ~ &~ $\eta$~ &~ $\Phi_2$  ~&~ $S$ ~&~ $S^-$ 
\\\hline 
%$SU(3)_C$ & $\bm{3}$  & $\bm{3}$  & $\bm{3}$ & $\bm{3}$ & $\bm{3}$  & $\bm{3}$  & $\bm{1}$  & $\bm{1}$   & $\bm{1}$  & $\bm{1} $  \\\hline 
 %%%
 $SU(2)_L$ & $\bm{2}$  & $\bm{1}$  & $\bm{1}$ & $\bm{2}$ & $\bm{1}$ & $\bm{1}$ &
 $\bm{2}$  & $\bm{1}$  & $\bm{2}$  & $\bm{2}$   & $\bm{1}$  & $\bm{1}$  \\\hline 
 %%%
$U(1)_Y$ & $-\frac12$ & $-1$  & $-1$ & $-\frac12$ & $0$ & $0$ &   $\frac{1}{2}$ & $0$
 & $\frac{1}{2}$ & $\frac12$  & $0$ &  $-1$   \\\hline
 %%%
 $U(1)_{R}$ & $0$ & $-x$  & $0$ & $0$ & $x$ & $0$  & $0$ & ${x}$  & $x$ & $-x$ & $0$  & $0$  \\\hline
 %%%
$Z_2$ & $+$ & $+$  & $+$ & $-$ & $-$ & $-$
& $+$ & $+$  & $-$ & $+$ & $-$ & $+$  \\\hline
%%%
%$\mathbb{Z}_2$ & $+$   & $-$  & $+$ & $+$& $-$& $-$& $+$ & $+$  \\\hline\hline
\end{tabular}
\caption{Field contents of fermions
and their charge assignments under $SU(2)_L\times U(1)_Y\times U(1)_R\times Z_2$, where each of the flavor index is defined as $\alpha\equiv 1-3$ and $i=1,2$.}
\label{tab:1}
% \end{tiny}
\end{table}
\end{center}
\end{widetext}

\section{Model setup}
%We discuss a two-loop induced radiative neutrino model. 
In this section, we review our model and derive formulas for neutrino mass matrix, lepton flavor violations, muon $g-2$, and relic density of DM. 
The particle contents and their charge assignments are shown in Tab.~\ref{tab:1}, in which only two inert bosons $\Phi_2$ and $S^-$ are introduced in addition to the previous work~\cite{Nomura:2016emz} to construct the Majorana type active neutrino masses at the two-loop level. Hence all the phenomenologies except neutrino sector can be retained. Notice here that the inert properties of $\Phi_2$ assure the local $U(1)_R$ symmetry even after spontaneous breaking via $\varphi$.
%; $\langle\varphi\rangle\equiv v'/\sqrt2$.  
%{\color{red} 
In addition, since $\varphi$ and $\Phi$ as well as all the SM fermions are $Z_2$ even, the $Z_2$ symmetry remains after the symmetry breaking where a neutral $Z_2$ odd particle can be a DM candidate.
Notice that we also need $N_R^i$ which is charged under $U(1)_R$ to cancel gauge anomaly. Here we assign odd $Z_2$ charge to $N_R^{i}$ in order to eliminate couplings that induce tree- and one-loop level neutrino mass generation in the previous model. In addition, we add $N_L^i$ to give Dirac mass  term like $M_D \bar N_L N_R$ which is generated by $\varphi \bar N_L N_R$ after $\varphi$ develops a VEV. 
Due to $Z_2$ symmetry $N_{L(R)}$ can be DM candidate, but we omit analysis for $N_{L(R)}$ DM since it is not related to neutrino mass generation.%}

Under these symmetries, the relevant Lagrangian and Higgs potential are given by
\begin{align}
-{\cal L}& \supset
(y_\ell)_{\alpha j}\bar L_\alpha \Phi_2 e_{R_j}  + (y'_\ell)_{ij} \bar L'_{ij}\eta e_{R_j} + (y_S)_{\alpha j}\bar L_\alpha L'_j S
 + (y'_S)_{\alpha \beta}\bar L^C_\alpha (i\sigma_2) L_\beta S^+\nn\\
&+(y_\tau)_{\alpha}\bar L_\alpha \tau_{R} (i\sigma_2)\Phi +m_{L'_k} \bar L'_k L'_k
-\lambda_1[\Phi^T(i\sigma_2)\Phi_2]\varphi S^-
-\lambda_2(\Phi^\dag\eta)\varphi^* S
+{\rm c.c.},
\label{eq:lag-quark}
\end{align}
where $\sigma_2$ is the Pauli matrix, $(\alpha,\beta)$ run over $1-3$, $(i,j,k)$ runs over $1,2$, and we abbreviate the full trivial potential. 
{Note that we have the Dirac mass term of $L'$ since it is introduced as vector-like. In addition, the term with coupling $y'_S$ is our source of lepton number violation in generating Majorana neutrino masses.}
Here we assume that the charged lepton mass is diagonal and we work on the basis where all the coefficients are real and positive for simplicity~\footnote{The first two columns, which comes from the first and second generations, are induced at the one-loop level. The third column, which corresponds to the third generation, are generated at the tree level.  See ref.~\cite{Nomura:2016emz} in details including scalar potential and $Z'$ boson associated with $U(1)_R$.}.

{Scalar bosons}:
In our model, scalar sector has five complex scalar fields $\{\Phi, \Phi_2, \eta, \varphi, S^\pm \}$ and one real singlet scalar $S$.
We parametrize the complex scalar fields as follows: 
\begin{align}
%\begin{tiny}
&\Phi =\left[
\begin{array}{c}
w^+\\
\frac{v+\phi+iz}{\sqrt2}
\end{array}\right],\quad 
%%%
\Phi_2 =\left[
\begin{array}{c}
\phi_2^+\\
{\phi_2 } 
\end{array}\right],\quad
%%%
\eta =\left[
\begin{array}{c}
\eta^+\\
\frac{\eta_{R}+i \eta_{I}} {\sqrt2}
\end{array}\right],\quad
\varphi\equiv \frac{v'+\varphi_R+i z_R}{\sqrt2},
\label{component}
%\end{tiny}
\end{align}
where $v(\approx 246$ GeV) and $v'$ are respectively vacuum expectation values (VEVs) of $\Phi$ and $\varphi$, and $w^\pm$, $z$, and $z_R$ are respectively Nambu-Goldstone (NG) bosons
which are absorbed by the longitudinal component of $W$, $Z$, and $Z'$ boson.
Since each of field set; $[\varphi,\phi]^T,\ [S,\eta_R]^T,[S^\pm,\phi_2^\pm]^T$, mixes through the terms of $|\Phi|^2|\varphi|^2$ and $\lambda_{1,2}$ after developing VEVs of $\Phi$ and $\varphi$, we define the relation between the mass eigenstate and the flavor eigenstate as follow:
\begin{align}
 &\left[\begin{array}{c} \varphi \\ \phi \end{array}\right] \equiv 
\left[\begin{array}{cc} \cos a & \sin a \\ -\sin a & \cos a \end{array}\right]
\left[\begin{array}{c} h_1 \\ h_2 \end{array}\right],\
%%%
\left[\begin{array}{c} S \\ \eta_R \end{array}\right] \equiv 
\left[\begin{array}{cc} \cos R & \sin R \\ -\sin R & \cos R \end{array}\right]
\left[\begin{array}{c} H_1 \\ H_2 \end{array}\right],\\\nn
%%%
&\left[\begin{array}{c} S^\pm \\ \phi_2^\pm \end{array}\right] \equiv 
\left[\begin{array}{cc} \cos C & \sin C \\ -\sin C & \cos C \end{array}\right]
\left[\begin{array}{c} H_1^\pm \\ H_2^\pm \end{array}\right],
\end{align}
where $h_2$ is the SM Higgs, and we hereafter write $\sin(\cos)$ as a short hand symbol; $\sin(\cos)\equiv s(c)$.

%%%%%%%%%%%%%%%%%%%
\begin{figure}[tbc]
\begin{center}
\includegraphics[scale=0.4]{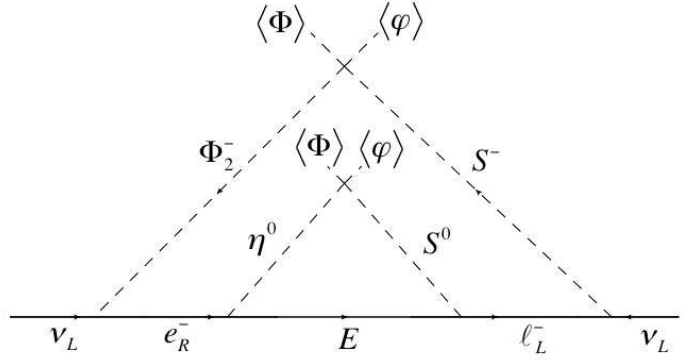}
%%%
\caption{The two loop diagram for generating Majorana mass term for active neutrinos.}
\label{loop}
\end{center}
\end{figure}
%%%%%%%%%%%%%%%%%%%
%\subsection{Neutrino mass matrix}
\subsection{ Neutrino mass matrix}
%%%
%{\color{red}
The dominant contribution to the active neutrino mass matrix $m_\nu$  is given at two-loop level where the corresponding diagram is shown in Fig.~\ref{loop}. 
Calculating the diagram, we obtain the formula of neutrino mass matrix such that %}
%\begin{widetext}
\begin{align}
(m_{\nu})_{ab}
&=-s_R c_R s_Cc_C\frac{(y_\ell)_{ai} (y'^\dag_\ell)_{ij} 
{\cal M}_{E_j} (y^\dag_S)_{jk} (y'^\dag_S)_{kb}+[(y_\ell)_{ai} (y'^\dag_\ell)_{ij} {\cal M}_{E_j} (y^\dag_S)_{jk} (y'^\dag_S)_{kb}]^T} 
{2(4\pi)^4},\\
{\cal M}_{E_j}&\equiv
m_{L'_j} \int[dx]\int[dX]\frac{y}{(x_4-1)^2}\ln\left(
\frac{\Delta_3[,E_j,H_1^\pm,H_2]\Delta_3[E_j,H_2^\pm,H_1] }
{\Delta_3[E_j,H_1^\pm,H_1] \Delta_3[E_j,H_2^\pm,H_2]}\right),\nn\\
\Delta_3[a,b,c]&\equiv
 x m_a^2-\frac{y (x_3 m^2_b + x_4 m^2_c )}{x_4^2-x_4},\nn\\
\int[dx]&\equiv \int_0^1 dx\int_0^{1-x} dy\delta(1-x-y),\nn\\
\int[dX]&\equiv\int_0^1dx_1\int_0^{1-x_1}dx_2 \int_0^{1-x_1-x_2}dx_3\int_0^{1-x_1-x_2-x_3} x_4\delta(1-x_1-x_2-x_3-x_4),
\end{align}%\end{widetext}
where we define $L'\equiv [E,N]^T$ and its mass is $m_{L'}$.

 % Remind here that two flavor of $E_k$(k=1-2) is introduced to obtain the current neutrino oscillation data.
The neutrino mass matrix $({m}_\nu)_{ab}$ can generally be diagonalized by the mixing matrix $V_{\rm MNS}$ (PMNS) and written in terms of experimental values depending on the normal hierarchy (NH) and inverted hierarchy (IH) as follows:
\begin{align}
({m}_\nu)^{exp.}_{ab} &=(V_{\rm MNS} D_\nu V_{\rm MNS}^T)_{ab},\quad D_\nu\equiv (m_{\nu_1},m_{\nu_2},m_{\nu_3})\nn\\
({\rm NH}):\
|{m}_\nu^{exp.}|&\approx \left[\begin{array}{ccc} 
0.0845-0.475 & 0.0629-0.971 &0.0411-0.964 \\
* & 1.44-3.49 &  1.94-2.85 \\
* & * &   1.22-3.33\\
  \end{array}\right]\times 10^{-11}\ {\rm GeV},\label{eq:exp-Neutmass-NH}
\\
({\rm IH}):\
|{m}_\nu^{exp.}|&\approx \left[\begin{array}{ccc} 
0.993-4.96 & 0.00261-3.83 &0.00280-3.95 \\
* & 0.00380-3.08 &  0.345-2.61 \\
* & * &   0.000647-3.30\\
  \end{array}\right]\times 10^{-11}\ {\rm GeV},\label{eq:exp-Neutmass-IH}
\\
V_{\rm MNS}&=
\left[\begin{array}{ccc} {c_{13}}c_{12} &c_{13}s_{12} & s_{13} e^{-i\delta}\\
 -c_{23}s_{12}-s_{23}s_{13}c_{12}e^{i\delta} & c_{23}c_{12}-s_{23}s_{13}s_{12}e^{i\delta} & s_{23}c_{13}\\
  s_{23}s_{12}-c_{23}s_{13}c_{12}e^{i\delta} & -s_{23}c_{12}-c_{23}s_{13}s_{12}e^{i\delta} & c_{23}c_{13}\\
  \end{array}\right]
  %%%
\left[\begin{array}{ccc} e^{i\alpha_1/2} & 0 &0 \\
0 & e^{i\alpha_2/2} & 0 \\
0 & 0 &  1\\
  \end{array}\right],
\end{align}
where we have used the following neutrino oscillation data at $3\sigma$~\cite{Forero:2014bxa} given by
\begin{align}
({\rm NH}):\ 
& 0.278 \leq s_{12}^2 \leq 0.375, \
 0.392 \leq s_{23}^2 \leq 0.643, \
 0.0177 \leq s_{13}^2 \leq 0.0294,  \  \delta\in [-\pi,\pi],
\nn \\
& 
 \ m_{\nu_3} =(\sqrt{23.0}-\sqrt{26.5}) \times10^{-11} \ {\rm GeV},  \
  \ m_{\nu_2} =(\sqrt{0.711}-\sqrt{0.818}) \times10^{-11} \ {\rm GeV}, \\
 %%%
({\rm IH}):\ 
& 0.278 \leq s_{12}^2 \leq 0.375, \
 0.403 \leq s_{23}^2 \leq 0.640, \
 0.0183 \leq s_{13}^2 \leq 0.0297,  \  \delta\in [-\pi,\pi],
  \nn\\
& 
%  m_{\nu_2} ({\rm eV}) = 0.0087,  \; 
 \ m_{\nu_1} =(\sqrt{22}-\sqrt{25.4}) \times10^{-11} \ {\rm GeV},  \
 % m_{\nu_3} ({\rm eV}) = 0.0502 .
  \ m_{\nu_2} =(\sqrt{22.711}-\sqrt{26.218}) \times10^{-11} \ {\rm GeV}, 
  \label{eq:neut-exp}
  \end{align}
and Majorana phases $\alpha_{1,2}$ taken to be $\alpha_{1,2}\in[-\pi,\pi]$ for both cases.
Notice here that we assume to be normal ordering to obtain the above numerical value of the neutrino mass matrix, and
we take one of three neutrino masses is zero, which is predicted by the theoretical aspect that $y'_S$ is anti-symmetric matrix,
therefore the rank of neutrino mass matrix is reduced to two. 
Applying the above nature, we can rewrite two components of $y'_S$ in terms of experimental values and one of the component of $y'_S$
as follows~\footnote{The detail analysis is found in ref.~\cite{Herrero-Garcia:2014hfa} for both hierarchies.}:
\begin{align}
({\rm NH}):\ (y'_S)_{e\tau}&=\left(\frac{s_{12}c_{23}}{c_{12}c_{13}} + \frac{s_{13} s_{23}}{c_{13}}e^{-i\delta} \right) (y'_S)_{\mu\tau},\quad 
(y'_S)_{e\mu}=\left(\frac{s_{12}c_{23}}{c_{12}c_{13}} - \frac{s_{13} s_{23}}{c_{13}}e^{-i\delta} \right)  (y'_S)_{\mu\tau},\label{eq:ys-NH}\\
%%%
({\rm IH}):\ (y'_S)_{e\tau}&=-\frac{s_{23}}{t_{13}} (y'_S)_{\mu\tau},\quad
(y'_S)_{e\mu}=\frac{c_{23}}{t_{13}} (y'_S)_{\mu\tau},\label{eq:ys-IH} 
  \end{align}
where the explicit form of $y'_S$ is given by
%%%%%%
  \begin{align}
(y'_S)&\equiv \left[\begin{array}{ccc}
0 & (y'_S)_{e\mu} & (y'_S)_{e\tau} \\
-(y'_S)_{e\mu}  & 0 & (y'_S)_{\mu\mu}  \\
-(y'_S)_{e\tau}  & -(y'_S)_{\mu\tau}  &  0\\
  \end{array}\right].
  \end{align}
In our numerical analysis, only $(y'_S)_{\mu\tau}$ is an input parameter, and we will search the allowed region of our parameter space by comparing the experimental values in Eqs.~(\ref{eq:exp-Neutmass-NH}) and ~(\ref{eq:exp-Neutmass-IH}).

\subsection{ Lepton Flavor Violations (LFVs) and muon anomalous magnetic moment}
{\it Lepton Flavor Violations}:
\label{lfv-lu}
The Yukawa couplings can induce LFV processes at loop level.
Here we focus on
$\ell_b\to\ell_a \gamma$ processes at one-loop level where $H_i$ and $E$ run inside a loop, and 
its branching ratio is written as
\begin{align}
{\rm BR}(\ell_i\to\ell_j \gamma)
=
\frac{48\pi^3 C_i\alpha_{\rm em}}{{\rm G_F^2} m_i^2 }(|(a_R)_{ij}|^2+|(a_L)_{ij}|^2),
\end{align}
where $\alpha_{\rm em}\approx1/137$ is the fine-structure constant,
$C_{b}=(1,1/5)$ for ($b=\mu,\tau$), ${\rm G_F}\approx1.17\times 10^{-5}$ GeV$^{-2}$ is the Fermi constant.
%%%
Calculating loop diagrams, $a_L$ and $a_R$ are respectively given as
\begin{align}
a_R&\equiv a_{R_0}+a_{R_1}+ a_{R_2},\quad
a_L\equiv a_{L_1}+ a_{L_2},\nn\\
%%%
(a_{R_0})_{ij}&=-\frac{s_R c_R}{2\sqrt2 (4\pi)^2}\sum_{k=1}^3
(y'_\ell)_{j,k} m_{L'_k} (y_S)_{k,i} \left(F_1[H_1,E_k]- F_1[H_2,E_k]\right),
\nn\\
%%%
(a_{R_1})_{ij}&=-\frac{1}{2\sqrt2 (4\pi)^2}\sum_{k=1}^3
\left[
\frac{2m_i (y_\ell)_{j,k} (y_\ell^\dag)_{k,i}}{m_{\phi_2}^2}
-
[m_j(y_\ell)^\dag_{j,k} (y_\ell)_{k,i}+m_i (y'_S)^\dag_{j,k} (y'_S)_{k,i}]
 \left( \frac{c_C^2}{m_{H_1}^\pm}+\frac{s_C^2}{m_{H_2}^\pm} \right)
 \right],
\nn\\
(a_{R_2})_{ij}&=\frac{1}{(4\pi)^2}\sum_{k=1}^3\left(
(y'_\ell)^\dag_{j,k} m_j (y'_\ell)_{k,i} F_2[E_k,\eta_\pm]-
(y_S)^\dag_{j,k} m_i (y_S)_{k,i} (c_R^2 F_2[H_1,E_k]+s_R^2 F_2[H_2,E_k]\right),
\nn\\
%%%
(a_{L_1})_{ij}&=-\frac{1}{2\sqrt2 (4\pi)^2}\sum_{k=1}^3
\left[
\frac{2m_j (y_\ell)_{j,k} (y_\ell^\dag)_{k,i}}{m_{\phi_2}^2}
-
[m_i(y_\ell)^\dag_{j,k} (y_\ell)_{k,i}+m_i (y'_S)^\dag_{j,k} (y'_S)_{k,i}]
 \left( \frac{c_C^2}{m_{H_1}^\pm}+\frac{s_C^2}{m_{H_2}^\pm} \right)
 \right],
\nn\\
(a_{L_2})_{ij}&=\frac{1}{(4\pi)^2}\sum_{k=1}^3\left(
(y'_\ell)^\dag_{j,k} m_i (y'_\ell)_{k,i} F_2[E_k,\eta_\pm]-
(y_S)^\dag_{j,k} m_j (y_S)_{k,i} (c_R^2 F_2[H_1,E_k]+s_R^2 F_2[H_2,E_k]\right),
\end{align} 
where 
\begin{align}
&F_1[a,b]\equiv
\frac{3 m_a^4 -4 m_a^2 m_b^2+m_b^4+2 m_a^4\ln(m_b^2/m_a^2)}{2(m_a^2-m_b^2)^3},\
F_1[a,a(=b)]\equiv
-\frac{1}{3 m_a^2}.
%\int_0^1dx\int_0^{1-x}dy\delta(1-x-y-z)\frac{x-1}{x m^2_a+(y+z)m^2_b},
\nn\\
%%%
&F_2[a,b]\equiv
\frac{2 m_a^6 +3 m_a^4 m_b^2 -6 m_a^2 m_b^4+ m_b^6 + 12 m_a^4 m_b^2 \ln(m_b/m_a)}{12(m_a^2-m_b^2)^4},
\
F_2[a,a(=b)]\equiv
\frac{1}{24m_a^2}.
%\int_0^1dx\int_0^{1-x}dy\delta(1-x-y-z)\frac{xy}{x m^2_a+(y+z)m^2_b}.
\end{align}
Here each of the experimental bound is summarized in table~\ref{tab:Cif}, and we impose these constraints in our numerical analysis.

\begin{table}[t]
\begin{tabular}{c|c|c|c} \hline
Process & $(i,j)$ & Experimental bounds ($90\%$ CL) & References \\ \hline
%%%%%%%
$\mu^{-} \to e^{-} \gamma$ & $(2,1)$ &
	${BR}(\mu \to e\gamma) < 4.2 \times 10^{-13}$ & \cite{TheMEG:2016wtm} \\
$\tau^{-} \to e^{-} \gamma$ & $(3,1)$ &
	${Br}(\tau \to e\gamma) < 3.3 \times 10^{-8}$ & \cite{Adam:2013mnn} \\
$\tau^{-} \to \mu^{-} \gamma$ & $(3,2)$ &
	${BR}(\tau \to \mu\gamma) < 4.4 \times 10^{-8}$ & \cite{Adam:2013mnn}   \\ \hline
%$\mu \to e\ {\rm conversion}$ & $(2,1)$ & ${R}(Ti) < 4.3 \times 10^{-12} \to{\cal O}(10^{-18})({\rm future\ bound })$ & \cite{Dohmen:1993mp}$\to$ \cite{Hungerford:2009zz, Cui:2009zz}   \\ \hline
\end{tabular}
\caption{Summary of $\ell_i \to \ell_j \gamma$ process and the upper bound of experimental data.}
\label{tab:Cif}
\end{table}

{\it Muon anomalous magnetic moment}:
%%%
Our formula of muon $g-2$ can be written in terms of $a_L$ and $a_R$, which have been derived in LFV sector as follows:
\begin{align}
\Delta a_\mu\approx -{m_\mu}(a_R+a_L)_{22},\label{damu}
\end{align}
where the lower index $2$ of $a_{R(L)}$ represents the muon.

\if0
The muon anomalous magnetic moment (muon $g-2$) has been 
measured at Brookhaven National Laboratory. 
The current average of the experimental results is given by~\cite{bennett}
\begin{align}
a^{\rm exp}_{\mu}=11 659 208.0(6.3)\times 10^{-10}. \notag
\end{align}
It has been well known that there is a discrepancy between the
experimental data and the prediction in the SM. 
The difference $\Delta a_{\mu}\equiv a^{\rm exp}_{\mu}-a^{\rm SM}_{\mu}$
was calculated in Ref.~\cite{discrepancy1} as 
\begin{align}
\Delta a_{\mu}=(29.0 \pm 9.0)\times 10^{-10}, \label{dev1}
\end{align}
and it was also derived in Ref.~\cite{discrepancy2} as
\begin{align}
\Delta a_{\mu}=(33.5 \pm 8.2)\times 10^{-10}. \label{dev2}
\end{align}
The above results given in Eqs. (\ref{dev1}) and (\ref{dev2}) correspond
to $3.2\sigma$ and $4.1\sigma$ deviations, respectively. 
\fi

%\newpage
\subsection{Dark Matter}
%{\it Dark Matter}:
Here we identify DM as $H_1(\equiv X)$ since it is correlated to neutrino mass matrix and LFVs. Then we provides formulas of nucleon-DM scattering cross section for the direct detection search and relic density which are taken into account in our numerical analysis later.
%%%

{\it Direct detection}:
We have a spin independent scattering cross section with nucleon through $h_{1,2}$ portal processes and its form is given by
\begin{align}
\sigma_N\approx 0.082\frac{m_N^4}{\pi v^2 M_X^2}\left(\frac{\mu_{2Xh_1}}{m_{h_1}^2}+\frac{\mu_{2Xh_2}}{m_{h_2}^2}\right)^2,
\end{align}
where $\mu_{2Xh_i}$ is the trilinear coupling of $X-X-h_i(i=1,2)$, and the mass of nucleon, which is symbolized by $m_N$, is around 0.939 GeV. Recent LUX experiment in 2016 reported the lower bound on $\sigma_N\lesssim$2.2$\times$ 10$^{-46}$ cm$^2$ at 50 GeV mass range of DM at  the 90 \% confidential level~\cite{Akerib:2016vxi}.
%%%

{\it Relic density}: Our relevant processes for the thermally averaged cross section comes from $2X\to 2h_2$, $2X\to f\bar f(f\approx top)$, and $2X\to VV^*(V=Z,W^\pm)$~\footnote{From the analysis of neutrino oscillation and LFVs, all the Yukawa couplings are so tiny that  the cross section coming from Yukawas are negligible.}.
%{\color{red} 
Here we note that when $s_R$ is not much small, $m_X \simeq m_{H_2} \simeq m_{\eta^\pm}$ is required for TeV scale DM due to large contribution from $VV^*$ channel without mass degeneration~\cite{Arhrib:2013ela}, and co-annihilation processes should be taken into account.
In our numerical analysis below, we take small $s_R$ so that wider range of $m_{H_2}$ and $m_{\eta^\pm}$ are allowed, and mass ranges outside co-annihilation region are discussed.%}
The squared amplitude for relevant processes and the approximated formula of relic density are given by~\cite{Griest:1990kh, Edsjo:1997bg}
\begin{align}
&\sigma v_{\rm rel}\approx \int_0^{\pi} \sin\theta d\theta \frac{ |\bar M|^2}{16\pi s}\sqrt{1-\frac{4 m_f^2}{s}},
\end{align}
where
%%%
\begin{align}
& |\bar M|^2\approx  |\bar M(2X\to 2h_2)|^2+ |\bar M(2X\to t\bar t)|^2+ |\bar M(2X\to VV^*)|^2,\\ 
%%%
&|\bar M(2X\to 2h_2)|^2
= \frac12
\left|\lambda_{2X2h_2}+\frac{\mu_{2Xh_1} \mu_{2h_2h_1} } {s-m_{h_1}^2}
+\frac{\mu_{2Xh_2} \mu_{2h_2h_2}} {s-m_{h_2}^2}\right.\nn\\
& \left.
+\mu_{Xh_1h_1}^2\left(\frac1{t-m_{h_1}^2}+\frac1{u-m_{h_1}^2}\right)
+\mu_{Xh_1h_2}^2\left(\frac1{t-m_{h_2}^2}+\frac1{u-m_{h_2}^2}\right)
 \right|^2,
\label{main-mode}
\\
%%%
&|\bar M(2X\to t\bar t)|^2
= 
6\left(\frac{m_t}{v}\right)^2 
\left|\frac{\mu_{2X h_1}}{s-M_{h_1}^2}+\frac{\mu_{2X h_2}}{s-M_{h_2}^2}\right|^2
(s-4m_t^2),\\
%%%
&|\bar M(2X\to2Z)|^2
= 
\frac{g_Z^4}{4} G^Z_{\mu\nu} G^Z_{\alpha\beta}\left(-g_{\mu\alpha}+\frac{k_{1\mu} k_{1\alpha}}{m_Z^2}\right)
\left(-g_{\nu\beta}+\frac{k_{2\nu} k_{2\beta}}{m_Z^2}\right),
\\
%%%
&|\bar M(2X\to W^+W^-)|^2
=
\frac{g_2^4}{4} G^W_{\mu\nu} G^W_{\alpha\beta}\left(-g_{\mu\alpha}+\frac{k_{1\mu} k_{1\alpha}}{m_W^2}\right)
\left(-g_{\nu\beta}+\frac{k_{2\nu} k_{2\beta}}{m_W^2}\right),\nn\\
%%%
&G^V_{ab}\equiv g_{ab}\left(s_R^2 - \frac{s_a v \mu_{2Xh_1}}{s-m^2_{h_1}}
+\frac{c_a v \mu_{2Xh_2}}{s-m^2_{h_2}} \right)\nn\\
%%%
&+\frac{s_R^2}{2}\left(\frac{(2p_1-k_1)_a (p_2-p_1+k_1)_b}{t-m^2_{\eta_V}}
+ \frac{(2p_1-k_2)_a (p_2-p_1+k_2)_b}{u-m^2_{\eta_V}}\right),
\end{align}
where $m_{\eta_Z}\equiv m_{\eta_I}$, $m_{\eta_W}\equiv m_{\eta^\pm}$, $p_{1,2}$ and $k_{1,2}$ are respectively the momentum of the initial and final state fields, and a trilinear coupling $\mu$ can be written in terms of quartic couplings and VEVs.
Then the relic density is given by
\begin{align}
\Omega h^2\approx \frac{1.07\times10^9 x_f^2}
{g^{1/2}_* M_{\rm pl} [{\rm GeV}] 
\left( a_{\rm eff}x_f+ 3 b_{\rm eff}  \right)},
\end{align}
where $g_*\approx 100$ is the total number of effective relativistic degrees of freedom at the time of freeze-out,
$M_{\rm pl}=1.22\times 10^{19}[{\rm GeV}] $ is Planck mass, $x_f\approx25$, and $ a_{\rm eff}$ and $ b_{\rm eff}$ are derived by
expanding $\sigma v_{\rm rel}$  in terms of $v_{\rm rel}$ up to $v_{\rm rel}^2$ as
\begin{align}
\sigma v_{\rm rel}\approx a_{\rm eff}+ b_{\rm eff} v^2_{\rm rel}.
\end{align}
The observed relic density reported by Planck suggest that $\Omega h^2\approx 0.12$~\cite{Ade:2013zuv}.

\section{Numerical results}
\begin{figure}[tbc]
\begin{center}
\includegraphics[scale=0.5]{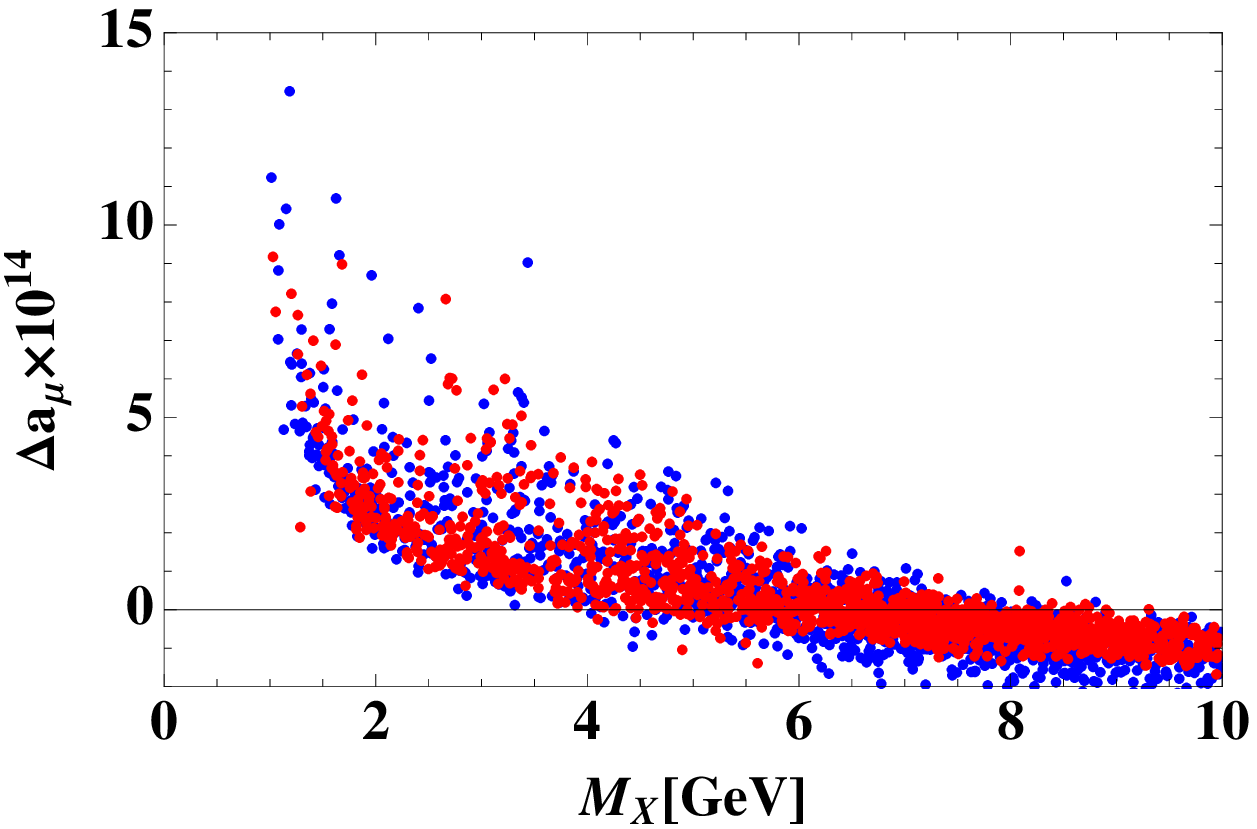}
\includegraphics[scale=0.5]{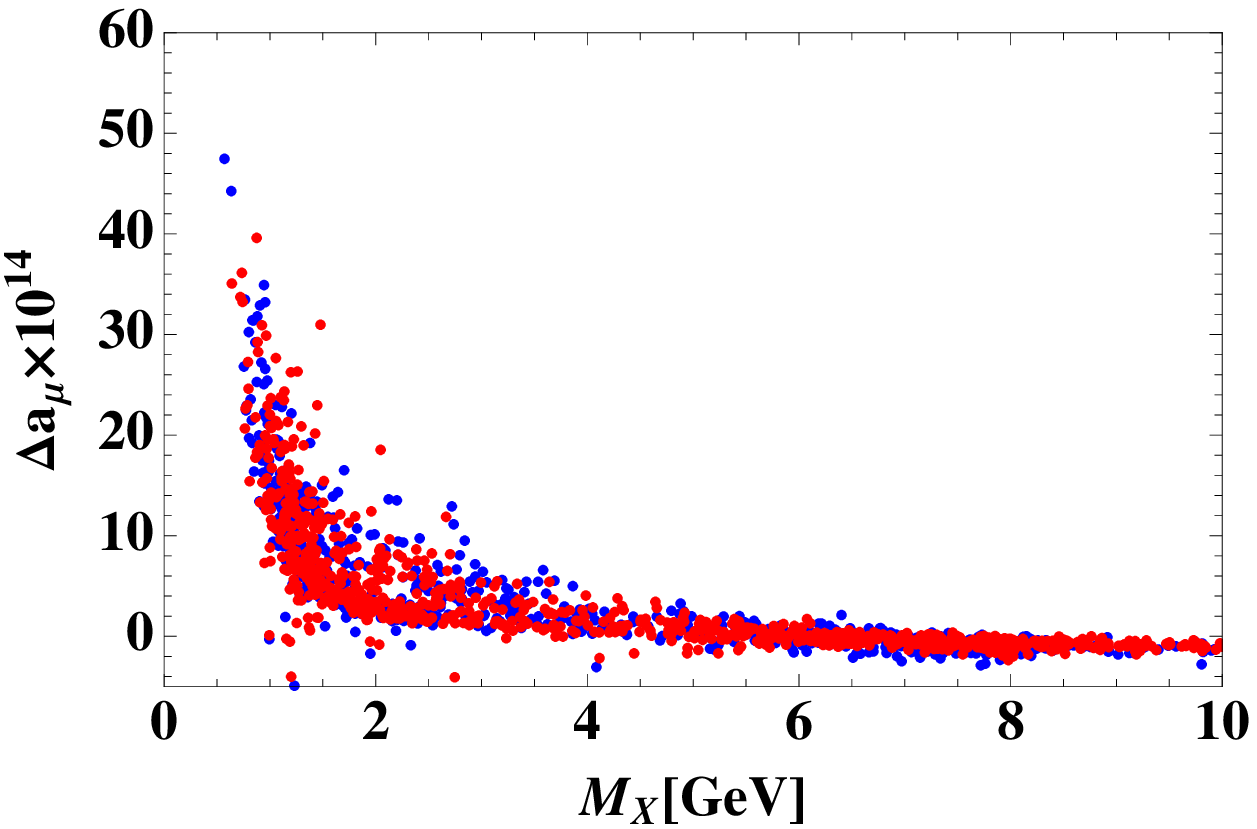}
%%%
\includegraphics[scale=0.5]{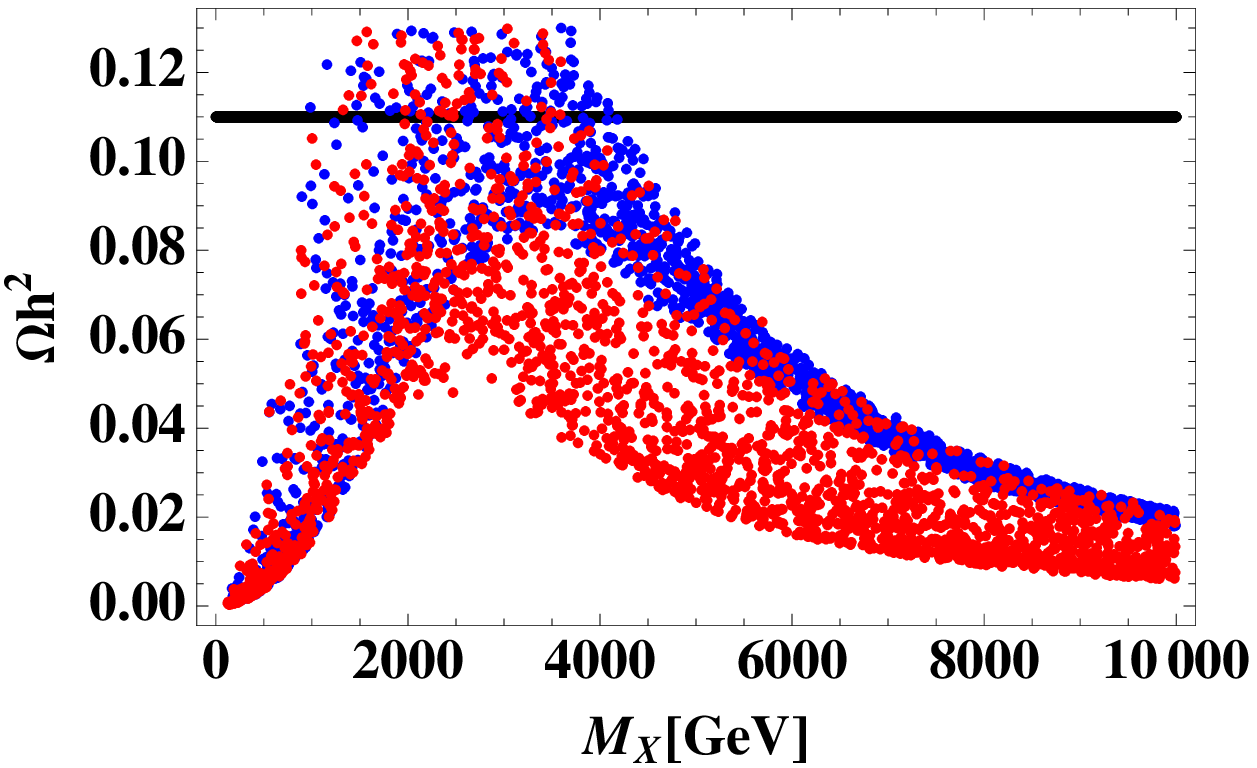}
\caption{The plots at the left-upper (NH) and the right-upper (IH) represent the allowed region in terms of the DM mass and the muon $g-2$. The red region is the case of $m_{\eta^\pm}=m_{H_2}\in[11 M_X/10, 12 M_X/10]$, and the maximum value of muon $g-2$ reaches ${\cal O}(1.0\times 10^{-13})$ for NH  and ${\cal O}(4.0\times 10^{-13})$ for IH.
On the other hand the blue one is the case of $m_{\eta^\pm}=m_{H_2}\in[11 M_X/10, 11.1 M_X/10]$, and the maximum value of muon $g-2$ reaches ${\cal O}(1.5\times10^{-13})$ for NH  and ${\cal O}(5.0\times 10^{-13})$ for IH.
One also find that the lower bound of the DM mass to be $1000 \, {\rm GeV}\lesssim M_X$, which comes from the LFVs for both cases.
%%%
The plot at the bottom represents the allowed region in terms of the DM mass and its relic density. The red region and blue region correspond to the same parameter setting as the upper figures. These analyses give the constraints on the DM mass and $\Delta a_\mu$; $1000\ {\rm GeV}\lesssim M_X\lesssim 4500$ GeV, and  $0\lesssim \Delta a_\mu\lesssim (1-1.5)\times 10^{-13}$ for NH and  $0\lesssim \Delta a_\mu\lesssim (4-5)\times 10^{-13}$ for IH.}
\label{num-figs}
\end{center}
\end{figure}

%First of all we fix $\lambda_{\Phi S}=0.05$ for simplicity in the numerical analysis.
%{\color{red} 
In this section, we carry out numerical analysis to search for the parameter region which can fit the experimental data.
To reduce number of free parameters, we first fix some parameters as follows; $s_a=0.3$,
$s_R= 0.1$, $s_C= 1$, % $s_R=\pi/36$, $s_C=\pi/4$
$m_{h_1}=4$ TeV.
We next randomly select values of masses as $M_X\in[0,10]$ TeV, $m_{\phi_2}\in[11M_X/10, 9]$ TeV, $m_{H_2^\pm}\in[11M_X/10,12 M_X/10]$ TeV, $\mu_{2Xh_i}=\mu_{2h_2h_i}=\mu_{Xh_2h_i}\equiv \mu\in[10^{-4},10^{-3}]$ GeV,~\footnote{{Notice here that the upper value $10^{-3}$ for $\mu_{2X h_i}$ is required to evade the constraint from the direct detection search, and other trilinear couplings are taken to be the same value for simplicity. When we relax the relation applying larger $\mu_{2h_2 h_i}$ and $\mu_{Xh_2 h_i}$, we have larger DM annihilation cross section reducing relic density without affecting neutrino mass matrix.  }} 
$m_{L'_1}\in[11M_X/10,9]$ TeV, $m_{L'_2}\in[m_{L'_1},9.5]$ TeV, {$m_{L'_3}\in[m_{L'_2},10]$} TeV, and $m_{\eta^\pm}=m_{H_2}\in[11 M_X/10, 12(11.1) M_X/10]$, where we consider the mass range of $m_{H_\eta^\pm}$ and $m_{H_2}$ outside co-annihilation region~\cite{Griest:1990kh} as discussed above.
Then, we randomly select values of the 29 parameters within $[-1,1]$ for all the dimensionless couplings to search for allowed parameter region.~\footnote{Totally we have 45(16 mass dimensions plus 29 dimensionless) input parameters, and 23 output parameters.}
As a result we find the allowed ranges for both NH and IH cases as follows: %}
\begin{align}
%& M_X \in [1\,\text{TeV}, 10\,\text{TeV}],\quad (M_E,\ M_N,\ m_{H_1},\ m_{\Delta^\pm},\ m_{\Delta^{\pm\pm}}) \in [M_X,\ 5\,\text{TeV}],\nn\\& \delta \in [0, 2\pi],\quad \alpha \in [-0.3, 0.3],\quad
(y'_S)_{\mu\tau}& \in [-0.027,0.027],\quad \lambda_{2X2h_2}\in[0.1,1], \\
%%%
y_S&= \left[\begin{array}{ccc} 
0.01-0.135 & -(0.025-0.02) &-(0.0265-0.02) \\
0.03-0.036 & 0.01-0.015 &  -(0.045-0.04) \\
-(0.027-0.02) & -(0.0208-0.02) &  -( 0.028-0.02)\\
  \end{array}\right],\nn\\
%%%
y_\ell&= \left[\begin{array}{ccc} 
-(0.016-0.01) & 0.04-0.0453 &-(0.02-0.01) \\
-(0.0063-0.001) & 0.04-0.05 &  -(0.031-0.01) \\
0.001-0.0012 & -(0.0174-0.01) &   0.03-0.036\\
  \end{array}\right],\nn\\
  %%%
  y'_\ell&= \left[\begin{array}{ccc} 
0.03-0.037 & -(0.0445-0.04) &0.001-0.0088 \\
0.01-0.0123 & 0.02-0.0275 &  -(0.032-0.01) \\
0.01-0.03 & 0.001-0.0094) &   0.001-0.005\\
  \end{array}\right],
	\label{range_scanning}
\end{align}
%%%
which can reproduce neutrino oscillation data, satisfies the constraints from LFVs and the direct detection searches~\footnote{We conservatively take the constraint $\sigma_N\lesssim 10^{-45}$cm$^2$ for all the mass region of DM.} and can provide the observed relic density of DM;  0.11$\le\Omega h^2\le$0.13. Here $\lambda_{2X2h_2}$ is the coupling constants of four point interaction $X$-$X$-$h_2$-$h_2$.
%, we fix some parameters as $s_a=0.3$, $s_R=\pi/36$, $s_C=\pi/4$, $m_{h_1}=4$ TeV, and we randomly select values of masses as
%$M_X\in[0,10]$ TeV, $m_{\phi_2}\in[11M_X/10, 9]$ TeV, $m_{H_2^\pm}\in[11M_X/10,12 M_X/10]$ TeV, $\mu_{2Xh_i}=\mu_{2hH_i}=\mu_{XhH_i}\equiv \mu\in[0.0001,0.001]$ GeV,
%$m_{L'_1}\in[11M_X/10,9]$ TeV, $m_{L'_2}\in[m_{L'_1},9.5]$ TeV , $m_{L'_2}\in[m_{L'_3},10]$ TeV, and$m_{\eta^\pm}=m_{H_2}\in[11 M_X/10, 12(11.1) M_X/10]$, where the lower bound $11 M_X/10$ is taken in order to evade any co-annihilation modes~\cite{Griest:1990kh}.
%{\color{red}
We thus find that the hierarchy of the required Yukawa coupling constants is not large. %}
For the parameter ranges, we generate $5\times 10^4$ million sample points and make figures to describe some correlations.
The plots at the left-upper (NH) and the right-upper (IH) in Fig.~\ref{num-figs} represent the allowed region in terms of the DM mass and the muon $g-2$. The red region is the case of $m_{\eta^\pm}=m_{H_2}\in[11 M_X/10, 12 M_X/10]$, and the maximum value of muon $g-2$ reaches ${\cal O}(1.0\times 10^{-13})$ for NH  and ${\cal O}(4.0\times 10^{-13})$ for IH.
On the other hands the blue one is the case of $m_{\eta^\pm}=m_{H_2}\in[11 M_X/10, 11.1 M_X/10]$, and the maximum value of muon $g-2$ reaches ${\cal O}(1.5\times 10^{-13})$ for NH  and ${\cal O}(5.0\times 10^{-13})$ for IH.
%%%
 One also find that the lower bound of the DM mass to be $1000 \, {\rm GeV}\lesssim M_X$, which comes from the LFVs for both cases. Notice here that the muon $g-2$ is much smaller than the current experimental value ${\cal O}(10^{-9})$~\cite{Bennett:2006fi}.

%%%%%%%%%%%%%%%%%%%
%%%%%%%%%%%%%%%%%%%

The plot at the bottom in Fig.~\ref{num-figs} represents the allowed region in terms of the DM mass and its relic density. The red region and blue region corresponds to the same parameter setting as the upper plots in Fig.~\ref{num-figs}. This analysis gives the definite constraint on the DM mass; $1000\ {\rm GeV}\lesssim M_X\lesssim 4500$ GeV. When the degeneracy between $M_X$ and $m_{\eta^\pm}$ relaxes, one finds that the bond of allowed region becomes to be wider. Notice here that it does not affect to the neutrino mass ordering because Yukawa couplings are neglected to the relic density.
For the lower range of the DM mass, the relic density decreases due to increasing annihilation cross section whose dominant mode is $2X\to 2h$. On the other hand, DM relic density decreases in the allowed regions with the larger DM mass, which comes from larger annihilation cross section for $2X\to VV^*$ mode;
it seems to violate the unitarity bound, however, this curve becomes flat in the limit $M_X\to \infty$.

\begin{figure}[tbc]
\begin{center}
\includegraphics[scale=0.5]{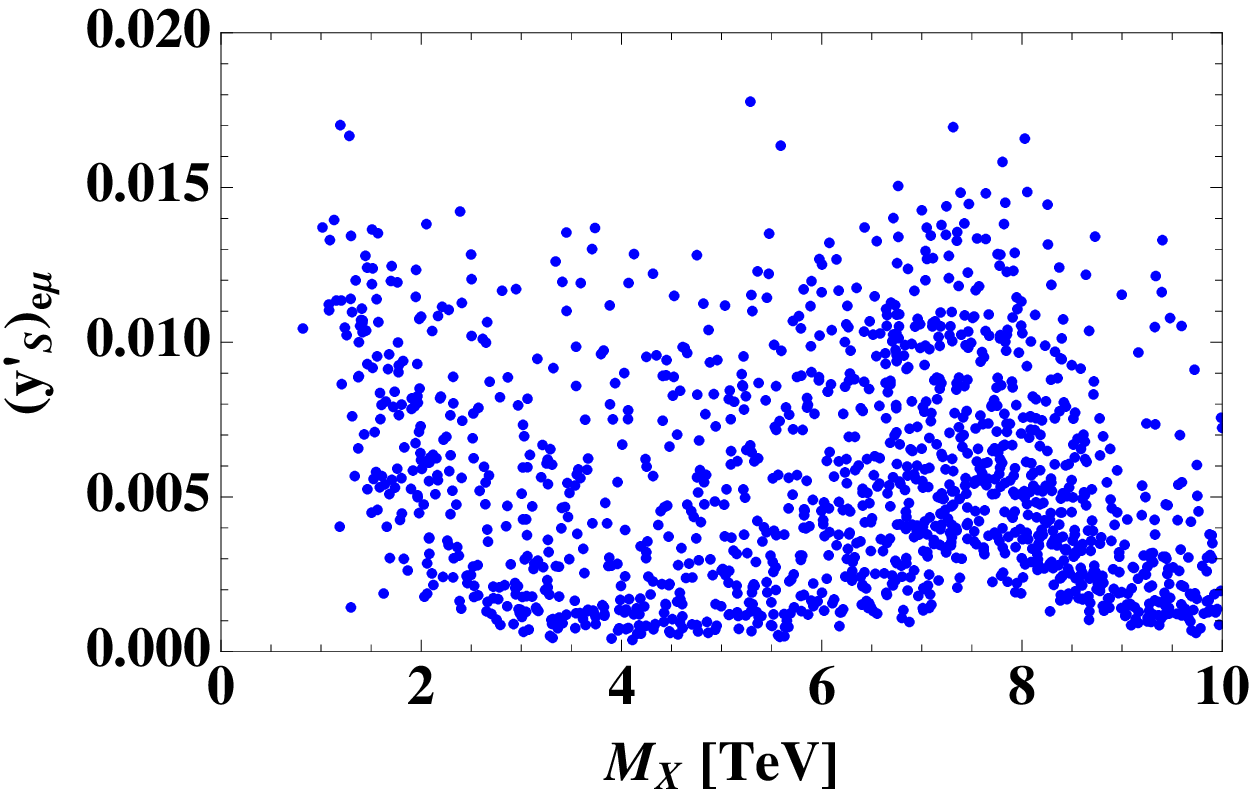}
\includegraphics[scale=0.5]{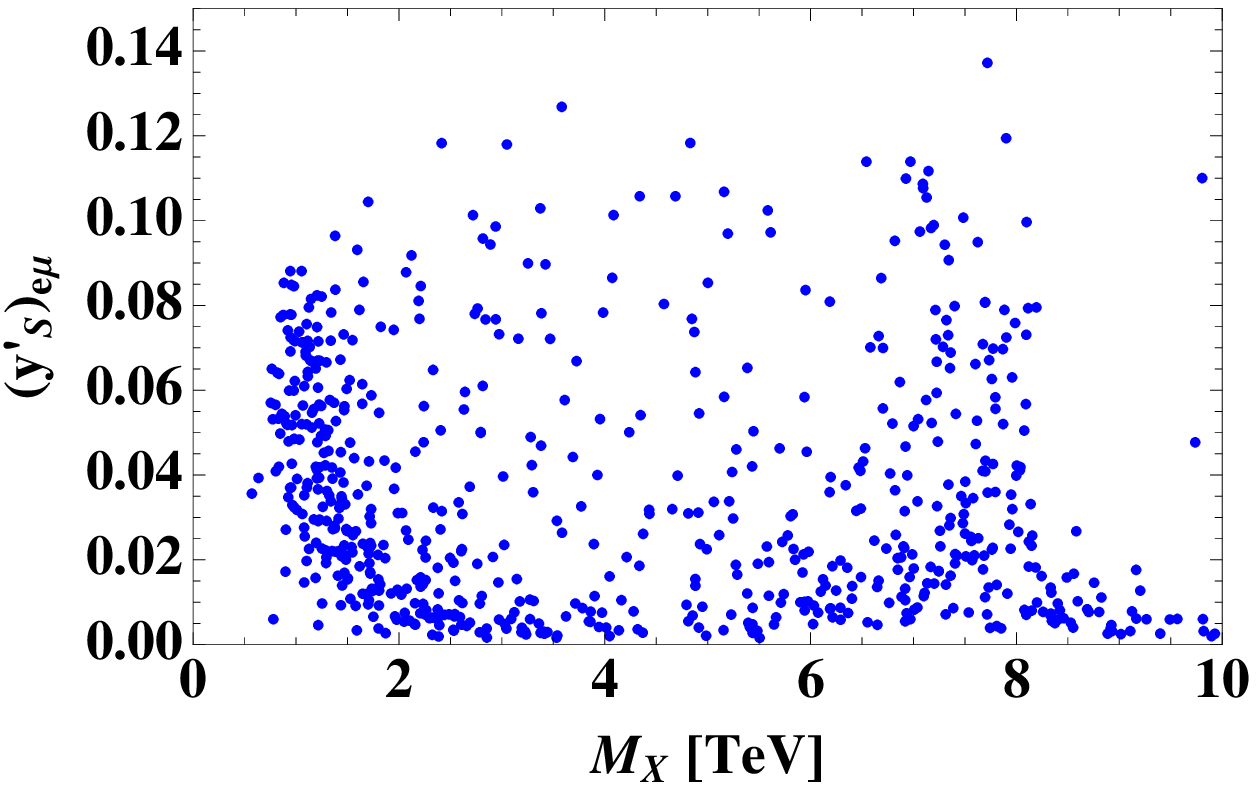}
%%%
\caption{The left figure (NH) and the right one (IH) represent the allowed region in terms of the DM mass and $(y'_S)_{e\mu}$. These figure tell us that its scale of IH is about ten times as large as the one of NH, where $m_{\eta^\pm}=m_{H_2}\in[11 M_X/10, 12 M_X/10]$ is used for both case, since its scale does not depend on the degeneracy. }
\label{ys12-figs}
\end{center}
\end{figure}

Accommodating the result of $1000\ {\rm GeV}\lesssim M_X\lesssim 4500$ GeV and feeding it back into the first figure, we also find the lower bound of muon $g-2$ that tells the negative value does not tend to be allowed.
In summary of our numerical analysis, we find as follows:
\begin{align}
&1000\ {\rm GeV}\lesssim M_X\lesssim 4500\ {\rm GeV},\\
({\rm NH}):\ 
&0\lesssim \Delta a_\mu \lesssim 1.0\times 10^{-13}\ {\rm for}\ m_{\eta^\pm}=m_{H_2}\in[11 M_X/10, 12 M_X/10],\\
&0\lesssim \Delta a_\mu \lesssim1.5\times 10^{-13}\ {\rm for}\ m_{\eta^\pm}=m_{H_2}\in[11 M_X/10, 11.1 M_X/10],\\
%%%
({\rm IH}):\ 
&0\lesssim \Delta a_\mu \lesssim 4.0\times 10^{-13}\ {\rm for}\ m_{\eta^\pm}=m_{H_2}\in[11 M_X/10, 12 M_X/10],\\
&0\lesssim \Delta a_\mu \lesssim 5.0\times 10^{-13}\ {\rm for}\ m_{\eta^\pm}=m_{H_2}\in[11 M_X/10, 11.1 M_X/10].
\end{align}
%%%
It is worthwhile mentioning that
lager value of muon $g-2$ tends to be obtained by the IH case. One of the main reason is that
the value of $(y'_S)_{e\mu}$, which contributes not to the constraint of $\mu\to e\gamma$ but to 
do the muon $g-2$, tends to be larger than the NH case as shown in Fig.~\ref{ys12-figs} due to the relation between $y'_S$ in Eqs.~(\ref{eq:ys-NH}) and (\ref{eq:ys-IH}).

%%%%%%%%%%%%%%%%%%%
%%%%%%%%%%%%%%%%%%%

%\section{Conclusions}
\section{ Conclusions and discussions}
We have studied a two-loop induced radiative neutrino model as an extension of our previous work in which the first and second generation standard model fermion masses in all the sector are induced at one-loop level.
Then we have discussed current neutrino oscillation data, lepton flavor violations, muon anomalous magnetic moment, and a bosonic dark matter candidate explaining the relic density under the direct detection constraint, considering both the normal and inverted neutrino mass hierarchy. 
%{\color{red} 
We have found that less hierarchical Yukawa couplings can fit the neutrino oscillation data satisfying the constrains from LFVs. %}
{Then we have found some results in terms of the DM mass, the muon $g-2$ ($\Delta a_\mu$) % and the key component of the anti-symmetric Yukawa coupling $y'_S$ where 
 where DM mass is in the range of 1 TeV to 4.5 TeV, and $0\lesssim \Delta a_\mu \lesssim 1.0(1.5) \times 10^{-13}$ with $m_{\eta^\pm}=m_{H_2}\in[11 M_X/10, 12(11.1) M_X/10]$ in NH case 
 while $0\lesssim \Delta a_\mu \lesssim 4.0(5.0)\times 10^{-13}$ with $m_{\eta^\pm}=m_{H_2}\in[11 M_X/10, 12(11.1) M_X/10]$ in IH case.
%\begin{align}
%&1000\ {\rm GeV}\lesssim M_X\lesssim 4500\ {\rm GeV},\\
%({\rm NH}):\ 
%&0\lesssim \Delta a_\mu \lesssim 1.0\times 10^{-13}\ {\rm for}\ m_{\eta^\pm}=m_{H_2}\in[11 M_X/10, 12 M_X/10],\\
%&0\lesssim \Delta a_\mu \lesssim1.5\times 10^{-13}\ {\rm for}\ m_{\eta^\pm}=m_{H_2}\in[11 M_X/10, 11.1 M_X/10],\\
%%%
%({\rm IH}):\ 
%&0\lesssim \Delta a_\mu \lesssim 4.0\times 10^{-13}\ {\rm for}\ m_{\eta^\pm}=m_{H_2}\in[11 M_X/10, 12 M_X/10],\\
%&0\lesssim \Delta a_\mu \lesssim 5.0\times 10^{-13}\ {\rm for}\ m_{\eta^\pm}=m_{H_2}\in[11 M_X/10, 11.1 M_X/10].
%\end{align}
%%%
In addition the key component of the anti-symmetric Yukawa coupling $y'_S$ can be order of $\sim 0.01(0.1)$ for NH(IH) cases.
It is worthwhile mentioning that
lager value of muon $g-2$ tends to be obtained by the IH case. One of the main reason is that
the value of $(y'_S)_{e\mu}$, which contributes not to the constraint of $\mu\to e\gamma$ but to 
do the muon $g-2$, tends to be larger than the NH case as shown in Fig.~\ref{ys12-figs} due to the relation between $y'_S$ in Eqs.~(\ref{eq:ys-NH}) and (\ref{eq:ys-IH}).
}
Especially we have found that IH is in favor of the muon $g-2$ although the maximum value is still smaller than the current experimental result.
%\section*{ Appendix}
%%%%%%%%%%%%%%%%%%%...

%\newpage
%%%%%%%%%%%%%%%%%%%%%%%%%%%%%%%%%%%
%\hspace{0.2cm} {\bf Acknowledgments}
%\section*{Acknowledgments}:
%\vspace{0.5cm}
\section*{Acknowledgments}
\vspace{0.5cm}
H. O. is sincerely grateful for all the KIAS members, Korean cordial persons, foods, culture, weather, and all the other things.
%%%%%%%%%%%%%%%%%%%%%%%%%%%%%%%%%%%
%%%%%%%%%%%%%%%%%%%%%%%%%%%%%%%%%%%


\begin{thebibliography}{99}


\bibitem{a-zee} 
A.~Zee,
 %``A Theory Of Lepton Number Violation, Neutrino Majorana Mass, And
 %Oscillation,''
 Phys.\ Lett.\  B {\bf 93}, 389 (1980)
 [Erratum-ibid.\  B {\bf 95}, 461  (1980)].

\bibitem{Cheng-Li} 
  T.~P.~Cheng and L.~F.~Li,
  %``Neutrino Masses, Mixings and Oscillations in SU(2) x U(1) Models of Electroweak Interactions,''
  Phys.\ Rev.\ D {\bf 22}, 2860 (1980).
  %%CITATION = PHRVA,D22,2860;%%

  %\cite{Pilaftsis:1991ug}
 \bibitem{Pilaftsis:1991ug} 
  A.~Pilaftsis,
  %``Radiatively induced neutrino masses and large Higgs neutrino couplings in the standard model with Majorana fields,''
  Z.\ Phys.\ C {\bf 55}, 275 (1992)
  doi:10.1007/BF01482590
  [hep-ph/9901206].
  %%CITATION = doi:10.1007/BF01482590;%%
  %231 citations counted in INSPIRE as of 25 Oct 2016

%\cite{Ma:2006km}
\bibitem{Ma:2006km} 
  E.~Ma,
  %``Verifiable radiative seesaw mechanism of neutrino mass and dark matter,''
  Phys.\ Rev.\ D {\bf 73}, 077301 (2006)
  [hep-ph/0601225].

  %\cite{Gu:2007ug}
\bibitem{Gu:2007ug} 
  P.~-H.~Gu and U.~Sarkar,
  %``Radiative Neutrino Mass, Dark Matter and Leptogenesis,''
  Phys.\ Rev.\ D {\bf 77}, 105031 (2008)
  [arXiv:0712.2933 [hep-ph]].
  %%CITATION = ARXIV:0712.2933;%%
  %10 citations counted in INSPIRE as of 20 Mar 2013

%\cite{Sahu:2008aw}
\bibitem{Sahu:2008aw} 
  N.~Sahu and U.~Sarkar;
  %``Extended Zee model for Neutrino Mass, Leptogenesis and Sterile Neutrino like Dark Matter,''
  Phys.\ Rev.\ D {\bf 78}, 115013 (2008)
  [arXiv:0804.2072 [hep-ph]].
  %%CITATION = ARXIV:0804.2072;%%
  %17 citations counted in INSPIRE as of 06 Jan 2015
  
%\cite{Gu:2008zf}
\bibitem{Gu:2008zf} 
  P.~-H.~Gu and U.~Sarkar,
  %``Radiative seesaw in left-right symmetric model,''
  Phys.\ Rev.\ D {\bf 78}, 073012 (2008)
  [arXiv:0807.0270 [hep-ph]].
  %%CITATION = ARXIV:0807.0270;%%
  %4 citations counted in INSPIRE as of 20 Mar 2013
  
%\cite{AristizabalSierra:2006ri}
\bibitem{AristizabalSierra:2006ri} 
  D.~Aristizabal Sierra and D.~Restrepo,
  %``Leptonic Charged Higgs Decays in the Zee Model,''
  JHEP {\bf 0608}, 036 (2006)
  [hep-ph/0604012].
  %%CITATION = HEP-PH/0604012;%%
  %15 citations counted in INSPIRE as of 13 juil. 2015
  
%\cite{Bouchand:2012dx}
\bibitem{Bouchand:2012dx} 
  R.~Bouchand and A.~Merle,
  %``Running of Radiative Neutrino Masses: The Scotogenic Model,''
  JHEP {\bf 1207}, 084 (2012)
  [arXiv:1205.0008 [hep-ph]].
  %%CITATION = ARXIV:1205.0008;%%
  %21 citations counted in INSPIRE as of 30 Jul 2014
  
%\cite{McDonald:2013hsa}
\bibitem{McDonald:2013hsa} 
  K.~L.~McDonald,
  %``Probing Exotic Fermions from a Seesaw/Radiative Model at the LHC,''
  JHEP {\bf 1311}, 131 (2013)
  [arXiv:1310.0609 [hep-ph]].
  %%CITATION = ARXIV:1310.0609;%%
  
%\cite{Ma:2014cfa}
\bibitem{Ma:2014cfa} 
  E.~Ma,
  %``Vanishing Higgs one-loop quadratic divergence in the scotogenic
%model and beyond,''
  Phys.\ Lett.\ B {\bf 732}, 167 (2014)
  [arXiv:1401.3284 [hep-ph]].
  %%CITATION = ARXIV:1401.3284;%%
  %4 citations counted in INSPIRE as of 30 Jul 2014

%\cite{Kajiyama:2013sza}
\bibitem{Kajiyama:2013sza} 
  Y.~Kajiyama, H.~Okada and K.~Yagyu,
  %``Electron/Muon Specific Two Higgs Doublet Model,''
  Nucl.\ Phys.\ B {\bf 887}, 358 (2014)
  [arXiv:1309.6234 [hep-ph]].
  %%CITATION = ARXIV:1309.6234;%%
  %20 citations counted in INSPIRE as of 22 mar 2015

%\cite{Kanemura:2011vm}
\bibitem{Kanemura:2011vm} 
  S.~Kanemura, O.~Seto and T.~Shimomura,
  %``Masses of dark matter and neutrino from TeV scale spontaneous
%$U(1)_{B-L}$ breaking,''
  Phys.\ Rev.\ D {\bf 84}, 016004 (2011)
  [arXiv:1101.5713 [hep-ph]].
  %%CITATION = ARXIV:1101.5713;%%

%\cite{Kanemura:2011jj}
\bibitem{Kanemura:2011jj}
 S.~Kanemura, T.~Nabeshima and H.~Sugiyama,
 %``Neutrino Masses from Loop-Induced Dirac Yukawa Couplings,''
 Phys.\ Lett.\ B {\bf 703}, 66 (2011)
 [arXiv:1106.2480 [hep-ph]].
 %%CITATION = ARXIV:1106.2480;%%

%\cite{Kanemura:2011mw}
\bibitem{Kanemura:2011mw} 
  S.~Kanemura, T.~Nabeshima and H.~Sugiyama,
  %``TeV-Scale Seesaw with Loop-Induced Dirac Mass Term and Dark Matter
%from $U(1)_{B-L}$ Gauge Symmetry Breaking,''
  Phys.\ Rev.\ D {\bf 85}, 033004 (2012)
  [arXiv:1111.0599 [hep-ph]].
  %%CITATION = ARXIV:1111.0599;%%

\bibitem{Schmidt:2012yg} 
  D.~Schmidt, T.~Schwetz and T.~Toma,
  %``Direct Detection of Leptophilic Dark Matter in a Model with
%Radiative Neutrino Masses,''
  Phys.\ Rev.\ D {\bf 85}, 073009 (2012)
  [arXiv:1201.0906 [hep-ph]].
  %%CITATION = ARXIV:1201.0906;%%

%\cite{Kanemura:2012rj}
\bibitem{Kanemura:2012rj} 
  S.~Kanemura and H.~Sugiyama,
  %``Dark matter and a suppression mechanism for neutrino masses in the Higgs triplet model,''
  Phys.\ Rev.\ D {\bf 86}, 073006 (2012)
  [arXiv:1202.5231 [hep-ph]].

%\cite{Farzan:2012sa}
\bibitem{Farzan:2012sa} 
  Y.~Farzan and E.~Ma,
  %``Dirac neutrino mass generation from dark matter,''
  Phys.\ Rev.\ D {\bf 86}, 033007 (2012)
  [arXiv:1204.4890 [hep-ph]].
  %%CITATION = ARXIV:1204.4890;%%
  %17 citations counted in INSPIRE as of 30 Jul 2014

%\cite{Kumericki:2012bf}
\bibitem{Kumericki:2012bf} 
  K.~Kumericki, I.~Picek and B.~Radovcic,
  %``Critique of Fermionic R\nuMDM and its Scalar Variants,''
  JHEP {\bf 1207}, 039 (2012)
  [arXiv:1204.6597 [hep-ph]].
  %%CITATION = ARXIV:1204.6597;%%
  %15 citations counted in INSPIRE as of 30 Jul 2014

%\cite{Kumericki:2012bh}
\bibitem{Kumericki:2012bh} 
  K.~Kumericki, I.~Picek and B.~Radovcic,
  %``TeV-scale Seesaw with Quintuplet Fermions,''
  Phys.\ Rev.\ D {\bf 86}, 013006 (2012)
  [arXiv:1204.6599 [hep-ph]].
  %%CITATION = ARXIV:1204.6599;%%
  %26 citations counted in INSPIRE as of 30 Jul 2014

%\cite{Ma:2012if}
\bibitem{Ma:2012if} 
  E.~Ma,
  %``Radiative Scaling Neutrino Mass and Warm Dark Matter,''
  Phys.\ Lett.\ B {\bf 717}, 235 (2012)
  [arXiv:1206.1812 [hep-ph]].
  %%CITATION = ARXIV:1206.1812;%%
  %21 citations counted in INSPIRE as of 30 Jul 2014

%\cite{Gil:2012ya}
\bibitem{Gil:2012ya} 
  G.~Gil, P.~Chankowski and M.~Krawczyk,
  %``Inert Dark Matter and Strong Electroweak Phase Transition,''
  Phys.\ Lett.\ B {\bf 717}, 396 (2012)
  [arXiv:1207.0084 [hep-ph]].
  %%CITATION = ARXIV:1207.0084;%%
  %22 citations counted in INSPIRE as of 30 Jul 2014

%\cite{Okada:2012np}
\bibitem{Okada:2012np} 
  H.~Okada and T.~Toma,
  %``Fermionic Dark Matter in Radiative Inverse Seesaw Model with U(1)_{B-L},''
  Phys.\ Rev.\ D {\bf 86}, 033011 (2012)
  arXiv:1207.0864 [hep-ph].
  %%CITATION = ARXIV:1207.0864;%%

%\cite{Hehn:2012kz}
\bibitem{Hehn:2012kz} 
  D.~Hehn and A.~Ibarra,
  %``A radiative model with a naturally mild neutrino mass hierarchy,''
  Phys.\ Lett.\ B {\bf 718}, 988 (2013)
  [arXiv:1208.3162 [hep-ph]].
  %%CITATION = ARXIV:1208.3162;%%
  %1 citations counted in INSPIRE as of 10 Mar 2013

%\cite{Dev:2012sg}
\bibitem{Dev:2012sg} 
  P.~S.~B.~Dev and A.~Pilaftsis,
  %``Minimal Radiative Neutrino Mass Mechanism for Inverse Seesaw Models,''
  Phys.\ Rev.\ D {\bf 86}, 113001 (2012)
  [arXiv:1209.4051 [hep-ph]].
  %%CITATION = ARXIV:1209.4051;%%
  %5 citations counted in INSPIRE as of 10 Mar 2013

%\cite{Kajiyama:2012xg}
\bibitem{Kajiyama:2012xg} 
  Y.~Kajiyama, H.~Okada and T.~Toma,
  %``Light Dark Matter Candidate in B-L Gauged Radiative Inverse
%Seesaw,''
  Eur.\ Phys.\ J.\ C {\bf 73}, 2381 (2013)
  [arXiv:1210.2305 [hep-ph]].
  %%CITATION = ARXIV:1210.2305;%%
  %13 citations counted in INSPIRE as of 30 Jul 2014

  %\cite{Toma:2013zsa}
\bibitem{Toma:2013zsa} 
  T.~Toma and A.~Vicente,
  %``Lepton Flavor Violation in the Scotogenic Model,''
  JHEP {\bf 1401}, 160 (2014)
%  doi:10.1007/JHEP01(2014)160
  [arXiv:1312.2840, arXiv:1312.2840 [hep-ph]].
  %%CITATION = doi:10.1007/JHEP01(2014)160;%%
  %32 citations counted in INSPIRE as of 31 Jul 2016

%\cite{Kanemura:2013qva}
\bibitem{Kanemura:2013qva} 
  S.~Kanemura, T.~Matsui and H.~Sugiyama,
  %``Loop Suppression of Dirac Neutrino Mass in the Neutrinophilic Two
%Higgs Doublet Model,''
  Phys.\ Lett.\ B {\bf 727}, 151 (2013)
  [arXiv:1305.4521 [hep-ph]].
  %%CITATION = ARXIV:1305.4521;%%
  %8 citations counted in INSPIRE as of 30 Jul 2014

%\cite{Law:2013saa}
\bibitem{Law:2013saa} 
  S.~S.~C.~Law and K.~L.~McDonald,
  %``A Class of Inert N-tuplet Models with Radiative Neutrino Mass and Dark Matter,''
  JHEP {\bf 1309}, 092 (2013)
  [arXiv:1305.6467 [hep-ph]].
  %%CITATION = doi:10.1007/JHEP09(2013)092;%%
  %27 citations counted in INSPIRE as of 21 Jan 2016

%\cite{Baek:2014qwa}
\bibitem{Baek:2014qwa} 
  S.~Baek and H.~Okada,
  %``7 keV Dark Matter as X-ray Line Signal in Radiative Neutrino Model,''
  arXiv:1403.1710 [hep-ph].
  %%CITATION = ARXIV:1403.1710;%%
  %45 citations counted in INSPIRE as of 08 May 2016

%\cite{Kanemura:2014rpa}
\bibitem{Kanemura:2014rpa} 
  S.~Kanemura, T.~Matsui and H.~Sugiyama,
  %``Neutrino Mass and Dark Matter from Gauged $U(1)_{B-L}$ Breaking,''
  Phys.\ Rev.\ D {\bf 90}, 013001 (2014)
  [arXiv:1405.1935 [hep-ph]].
  %%CITATION = ARXIV:1405.1935;%%

%\cite{Fraser:2014yha}
\bibitem{Fraser:2014yha} 
  S.~Fraser, E.~Ma and O.~Popov,
  %``Scotogenic Inverse Seesaw Model of Neutrino Mass,''
  Phys.\ Lett.\ B {\bf 737}, 280 (2014)
  [arXiv:1408.4785 [hep-ph]].
  %%CITATION = ARXIV:1408.4785;%%

  %\cite{Vicente:2014wga}
\bibitem{Vicente:2014wga} 
  A.~Vicente and C.~E.~Yaguna,
  %``Probing the scotogenic model with lepton flavor violating processes,''
  JHEP {\bf 1502}, 144 (2015)
%  doi:10.1007/JHEP02(2015)144
  [arXiv:1412.2545 [hep-ph]].
  %%CITATION = doi:10.1007/JHEP02(2015)144;%%
  %14 citations counted in INSPIRE as of 31 Jul 2016

%\cite{Baek:2015mna}
\bibitem{Baek:2015mna} 
  S.~Baek, H.~Okada and K.~Yagyu,
  %``Flavour Dependent Gauged Radiative Neutrino Mass Model,''
  JHEP {\bf 1504}, 049 (2015)
  [arXiv:1501.01530 [hep-ph]].
  %%CITATION = ARXIV:1501.01530;%%
  %7 citations counted in INSPIRE as of 26 juin 2015

  %\cite{Merle:2015gea}
\bibitem{Merle:2015gea} 
  A.~Merle and M.~Platscher,
  %``Parity Problem of the Scotogenic Neutrino Model,''
  Phys.\ Rev.\ D {\bf 92}, no. 9, 095002 (2015)
 % doi:10.1103/PhysRevD.92.095002
  [arXiv:1502.03098 [hep-ph]].
  %%CITATION = doi:10.1103/PhysRevD.92.095002;%%
  %10 citations counted in INSPIRE as of 31 Jul 2016

%\cite{Restrepo:2015ura}
\bibitem{Restrepo:2015ura} 
  D.~Restrepo, A.~Rivera, M.~S\'anchez-Pel\'aez, O.~Zapata and W.~Tangarife,
  %``Radiative neutrino masses in the singlet-doublet fermion dark matter model with scalar singlets,''
  arXiv:1504.07892 [hep-ph].
  %%CITATION = ARXIV:1504.07892;%%

  %\cite{Merle:2015ica}
\bibitem{Merle:2015ica} 
  A.~Merle and M.~Platscher,
  %``Running of radiative neutrino masses: the scotogenic model — revisited,''
  JHEP {\bf 1511}, 148 (2015)
%  doi:10.1007/JHEP11(2015)148
  [arXiv:1507.06314 [hep-ph]].
  %%CITATION = doi:10.1007/JHEP11(2015)148;%%
  %5 citations counted in INSPIRE as of 31 Jul 2016

%\cite{Wang:2015saa}
\bibitem{Wang:2015saa} 
  W.~Wang and Z.~L.~Han,
  %``Radiative linear seesaw model, dark matter, and $U(1)_{B-L}$,''
  Phys.\ Rev.\ D {\bf 92}, 095001 (2015)
  [arXiv:1508.00706 [hep-ph]].
  %%CITATION = doi:10.1103/PhysRevD.92.095001;%%
  %5 citations counted in INSPIRE as of 21 Jan 2016

 \bibitem{Ahn:2012cg} 
  Y.~H.~Ahn and H.~Okada,
  %``Non-zero $\theta_{13}$ linking to Dark Matter from Non-Abelian
%Discrete Flavor Model in Radiative Seesaw,''
  Phys.\ Rev.\ D {\bf 85}, 073010 (2012)
  [arXiv:1201.4436 [hep-ph]].
  %%CITATION = ARXIV:1201.4436;%%

%\cite{Ma:2012ez}
\bibitem{Ma:2012ez} 
  E.~Ma, A.~Natale and A.~Rashed,
  %``Scotogenic $A_4$ Neutrino Model for Nonzero $\theta_{13}$ and Large
%$\delta_{CP}$,''
  Int.\ J.\ Mod.\ Phys.\ A {\bf 27}, 1250134 (2012)
  [arXiv:1206.1570 [hep-ph]].
  %%CITATION = ARXIV:1206.1570;%%

%\cite{Hernandez:2013dta}
\bibitem{Hernandez:2013dta} 
  A.~E.~Carcamo Hernandez, I.~d.~M.~Varzielas, S.~G.~Kovalenko, H.~P\"{a}s and I.~Schmidt,
  %``Lepton masses and mixings in a $A_{4}$ multi-Higgs model with radiative seesaw mechanism,''
  Phys.\ Rev.\ D {\bf 88}, 076014 (2013)
  [arXiv:1307.6499 [hep-ph]].
  %%CITATION = ARXIV:1307.6499;%%
  %2 citations counted in INSPIRE as of 25 Nov 2013

%\cite{Ma:2014eka}
\bibitem{Ma:2014eka} 
  E.~Ma and A.~Natale,
  %``Scotogenic $Z_2$ or $U(1)_D$ Model of Neutrino Mass with
%$\Delta(27)$ Symmetry,''
  Phys.\ Lett.\ B {\bf 723}, 403 (2014)
  [arXiv:1403.6772 [hep-ph]].
  %%CITATION = ARXIV:1403.6772;%%

%\cite{Ma:2014yka}
\bibitem{Ma:2014yka} 
  E.~Ma,
  %``Syndetic Model of Fundamental Interactions,''
  Phys.\ Lett.\ B {\bf 741}, 202 (2015)
  [arXiv:1411.6679 [hep-ph]].
  %%CITATION = ARXIV:1411.6679;%%
  %4 citations counted in INSPIRE as of 22 mar 2015

  %\cite{Ma:2015pma}
\bibitem{Ma:2015pma} 
  E.~Ma,
  %``Transformative A_4 Mixing of Neutrinos with CP Violation,''
  arXiv:1504.02086 [hep-ph].
  %%CITATION = ARXIV:1504.02086;%%
  %4 citations counted in INSPIRE as of 26 Jun 2015

  %\cite{Ma:2013mga}
\bibitem{Ma:2013mga} 
  E.~Ma,
  %``Radiative Origin of All Quark and Lepton Masses through Dark Matter
%with Flavor Symmetry,''
  Phys.\ Rev.\ Lett.\  {\bf 112}, 091801 (2014)
  [arXiv:1311.3213 [hep-ph]].
  %%CITATION = ARXIV:1311.3213;%%

  %\cite{Okada:2013iba}
\bibitem{radlepton1} 
  H.~Okada and K.~Yagyu,
  %``Radiative Generation of the Lepton Mass,''
  Phys.\ Rev.\ D {\bf 89}, 053008 (2014)
  [arXiv:1311.4360 [hep-ph]].
  %%CITATION = ARXIV:1311.4360;%%
  %8 citations counted in INSPIRE as of 30 Jul 2014

%\cite{Okada:2014nsa}
\bibitem{Okada:2014nsa} 
  H.~Okada and K.~Yagyu;
  %``Radiative generation of lepton masses with the U(1)′ gauge symmetry,''
  Phys.\ Rev.\ D {\bf 90}, no. 3, 035019 (2014)
  [arXiv:1405.2368 [hep-ph]].
  %%CITATION = ARXIV:1405.2368;%%
  %11 citations counted in INSPIRE as of 22 Mar 2015

  %\cite{Brdar:2013iea}
\bibitem{Brdar:2013iea} 
  V.~Brdar, I.~Picek and B.~Radovcic,
  %``Radiative Neutrino Mass with Scotogenic Scalar Triplet,''
  Phys.\ Lett.\ B {\bf 728}, 198 (2014)
  [arXiv:1310.3183 [hep-ph]].
  %%CITATION = ARXIV:1310.3183;%%
  %13 citations counted in INSPIRE as of 08 sept. 2015

%\cite{Okada:2015kkj}
\bibitem{Okada:2015kkj} 
  H.~Okada, Y.~Orikasa and T.~Toma,
  %``Non-thermal Dark Matter Models and Signals,''
  arXiv:1511.01018 [hep-ph].
  %%CITATION = ARXIV:1511.01018;%%

%\cite{Bonnet:2012kz}
\bibitem{Bonnet:2012kz} 
  F.~Bonnet, M.~Hirsch, T.~Ota and W.~Winter,
  %``Systematic study of the d=5 Weinberg operator at one-loop order,''
  JHEP {\bf 1207}, 153 (2012)
  [arXiv:1204.5862 [hep-ph]].
  %%CITATION = ARXIV:1204.5862;%%
  %32 citations counted in INSPIRE as of 30 Jul 2014

  %\cite{Joaquim:2014gba}
\bibitem{Joaquim:2014gba} 
  F.~R.~Joaquim and J.~T.~Penedo,
  %``Radiative charged-lepton mass generation in multi-Higgs doublet models,''
  Phys.\ Rev.\ D {\bf 90}, no. 3, 033011 (2014)
%  doi:10.1103/PhysRevD.90.033011
  [arXiv:1403.4925 [hep-ph]].
  %%CITATION = doi:10.1103/PhysRevD.90.033011;%%
  %6 citations counted in INSPIRE as of 31 Jul 2016

%\cite{Davoudiasl:2014pya}
\bibitem{Davoudiasl:2014pya} 
  H.~Davoudiasl and I.~M.~Lewis,
  %``Right-Handed Neutrinos as the Origin of the Electroweak Scale,''
  Phys.\ Rev.\ D {\bf 90}, no. 3, 033003 (2014)
  [arXiv:1404.6260 [hep-ph]].
  %%CITATION = ARXIV:1404.6260;%%
  %9 citations counted in INSPIRE as of 22 Mar 2015

  %\cite{Lindner:2014oea}
\bibitem{Lindner:2014oea} 
  M.~Lindner, S.~Schmidt and J.~Smirnov,
  %``Neutrino Masses and Conformal Electro-Weak Symmetry Breaking,''
  arXiv:1405.6204 [hep-ph];
  %%CITATION = ARXIV:1405.6204;%%

%\cite{Okada:2014nea}
\bibitem{Okada:2014nea} 
  H.~Okada and Y.~Orikasa,
  %``Classically conformal radiative neutrino model with gauged B − L symmetry,''
  Phys.\ Lett.\ B {\bf 760}, 558 (2016)
  doi:10.1016/j.physletb.2016.07.039
  [arXiv:1412.3616 [hep-ph]].
  %%CITATION = doi:10.1016/j.physletb.2016.07.039;%%
  %39 citations counted in INSPIRE as of 29 Nov 2016


%\cite{Mambrini:2015sia}
 \bibitem{Mambrini:2015sia} 
  Y.~Mambrini, S.~Profumo and F.~S.~Queiroz,
  %``Dark Matter and Global Symmetries,''
  arXiv:1508.06635 [hep-ph].
  %%CITATION = ARXIV:1508.06635;%%

  %\cite{Boucenna:2014zba}
\bibitem{Boucenna:2014zba} 
  S.~M.~Boucenna, S.~Morisi and J.~W.~F.~Valle,
  %``The low-scale approach to neutrino masses,''
  Adv.\ High Energy Phys.\  {\bf 2014}, 831598 (2014)
%  doi:10.1155/2014/831598
  [arXiv:1404.3751 [hep-ph]].
  %%CITATION = doi:10.1155/2014/831598;%%
  %29 citations counted in INSPIRE as of 01 Mar 2016

  %\cite{Ahriche:2016acx}
\bibitem{Ahriche:2016acx} 
  A.~Ahriche, S.~M.~Boucenna and S.~Nasri,
  %``Dark Radiative Inverse Seesaw,''
  arXiv:1601.04336 [hep-ph].
  %%CITATION = ARXIV:1601.04336;%%

%\cite{Fraser:2015mhb}
\bibitem{Fraser:2015mhb} 
  S.~Fraser, C.~Kownacki, E.~Ma and O.~Popov,
  %``Type II Radiative Seesaw Model of Neutrino Mass with Dark Matter,''
  arXiv:1511.06375 [hep-ph].
  %%CITATION = ARXIV:1511.06375;%% 

  %\cite{Fraser:2015zed}
\bibitem{Fraser:2015zed} 
  S.~Fraser, E.~Ma and M.~Zakeri,
  %``Verifiable Associated Processes from Radiative Lepton Masses with Dark Matter,''
  arXiv:1511.07458 [hep-ph].
  %%CITATION = ARXIV:1511.07458;%%
  %1 citations counted in INSPIRE as of 19 Dec 2015

  %\cite{Adhikari:2015woo}
\bibitem{Adhikari:2015woo} 
  R.~Adhikari, D.~Borah and E.~Ma,
  %``New U(1) Gauge Model of Radiative Lepton Masses with Sterile Neutrino and Dark Matter,''
  arXiv:1512.05491 [hep-ph].
  %%CITATION = ARXIV:1512.05491;%%

%\cite{Okada:2015vwh}
\bibitem{Okada:2015vwh} 
  H.~Okada and Y.~Orikasa,
  %``Radiative neutrino model with an inert triplet scalar,''
  Phys.\ Rev.\ D {\bf 94}, no. 5, 055002 (2016)
  doi:10.1103/PhysRevD.94.055002
  [arXiv:1512.06687 [hep-ph]].
  %%CITATION = doi:10.1103/PhysRevD.94.055002;%%
  %13 citations counted in INSPIRE as of 29 Nov 2016
  
  
  
  %\cite{Ibarra:2016dlb}
\bibitem{Ibarra:2016dlb} 
  A.~Ibarra, C.~E.~Yaguna and O.~Zapata,
  %``Direct Detection of Fermion Dark Matter in the Radiative Seesaw Model,''
  Phys.\ Rev.\ D {\bf 93}, no. 3, 035012 (2016)
%  doi:10.1103/PhysRevD.93.035012
  [arXiv:1601.01163 [hep-ph]].
  %%CITATION = doi:10.1103/PhysRevD.93.035012;%%
  %7 citations counted in INSPIRE as of 31 Jul 2016

  %\cite{Arbelaez:2016mhg}
\bibitem{Arbelaez:2016mhg} 
  C.~Arbelaez, A.~E.~C.~Hernandez, S.~Kovalenko and I.~Schmidt,
  %``Linking radiative seesaw-type mechanism of fermion masses and non-trivial quark mixing with the 750 GeV diphoton excess,''
  arXiv:1602.03607 [hep-ph].
  %%CITATION = ARXIV:1602.03607;%%
  %19 citations counted in INSPIRE as of 28 Jun 2016

%\cite{Ahriche:2016rgf}
\bibitem{Ahriche:2016rgf} 
  A.~Ahriche, K.~L.~McDonald, S.~Nasri and I.~Picek,
  %``A Critical Analysis of One-Loop Neutrino Mass Models with Minimal Dark Matter,’'
  Phys.\ Lett.\ B {\bf 757}, 399 (2016)
 % doi:10.1016/j.physletb.2016.04.022
  [arXiv:1603.01247 [hep-ph]].
  %%CITATION = doi:10.1016/j.physletb.2016.04.022;%%
  %10 citations counted in INSPIRE as of 31 Jul 2016

  %\cite{Lu:2016ucn}
\bibitem{Lu:2016ucn} 
  W.~B.~Lu and P.~H.~Gu,
  %``Leptogenesis, radiative neutrino masses and inert Higgs triplet dark matter,''
  arXiv:1603.05074 [hep-ph].
  %%CITATION = ARXIV:1603.05074;%%
  %3 citations counted in INSPIRE as of 08 May 2016

    %\cite{Kownacki:2016hpm}
\bibitem{Kownacki:2016hpm} 
  C.~Kownacki and E.~Ma,
  %``Gauge $U(1)$ Dark Symmetry and Radiative Light Fermion Masses,''
  arXiv:1604.01148 [hep-ph].
  %%CITATION = ARXIV:1604.01148;%%

  %\cite{Ahriche:2016cio}
\bibitem{Ahriche:2016cio} 
  A.~Ahriche, K.~L.~McDonald and S.~Nasri,
  %``The Scale-Invariant Scotogenic Model,''
  arXiv:1604.05569 [hep-ph].
  %%CITATION = ARXIV:1604.05569;%%
  %1 citations counted in INSPIRE as of 08 May 2016

  %\cite{Ahriche:2016ixu}
\bibitem{Ahriche:2016ixu} 
  A.~Ahriche, A.~Manning, K.~L.~McDonald and S.~Nasri,
  %``Scale-Invariant Models with One-Loop Neutrino Mass and Dark Matter Candidates,''
  arXiv:1604.05995 [hep-ph].
  %%CITATION = ARXIV:1604.05995;%%

    %\cite{Ma:2016nnn}
\bibitem{Ma:2016nnn} 
  E.~Ma, N.~Pollard, O.~Popov and M.~Zakeri,
  %``Gauge $B-L$ Model of Radiative Neutrino Mass with Multipartite Dark Matter,''
  arXiv:1605.00991 [hep-ph].
  %%CITATION = ARXIV:1605.00991;%%

%\cite{Nomura:2016jnl}
\bibitem{Nomura:2016jnl} 
  T.~Nomura, H.~Okada and Y.~Orikasa,
  %``$SU(2)_L$ septet scalar linking to a radiative neutrino model,''
  Phys.\ Rev.\ D {\bf 94}, no. 5, 055012 (2016)
  doi:10.1103/PhysRevD.94.055012
  [arXiv:1605.02601 [hep-ph]].
  %%CITATION = doi:10.1103/PhysRevD.94.055012;%%
  %8 citations counted in INSPIRE as of 29 Nov 2016

 
  %\cite{Hagedorn:2016dze}
 \bibitem{Hagedorn:2016dze} 
  C.~Hagedorn, T.~Ohlsson, S.~Riad and M.~A.~Schmidt,
  %``Unification of Gauge Couplings in Radiative Neutrino Mass Models,''
  arXiv:1605.03986 [hep-ph].
  %%CITATION = ARXIV:1605.03986;%%
  %1 citations counted in INSPIRE as of 25 Jun 2016

  %\cite{Antipin:2016awv}
\bibitem{Antipin:2016awv} 
  O.~Antipin, P.~Culjak, K.~Kumericki and I.~Picek,
  %``Radiative neutrino models in light of diphoton signals,''
  arXiv:1606.05163 [hep-ph].
  %%CITATION = ARXIV:1606.05163;%%


%\cite{Nomura:2016emz}
  \bibitem{Nomura:2016emz} 
  T.~Nomura and H.~Okada,
  %``Radiatively induced Quark and Lepton Mass Model,''
  Phys.\ Lett.\ B {\bf 761}, 190 (2016)
 % doi:10.1016/j.physletb.2016.08.023
  [arXiv:1606.09055 [hep-ph]].
   %%CITATION = doi:10.1016/j.physletb.2016.08.023;%%
  %3 citations counted in INSPIRE as of 03 Sep 2016
 
 %\cite{Gu:2016ghu}
 \bibitem{Gu:2016ghu} 
  P.~H.~Gu, E.~Ma and U.~Sarkar,
  %``Connecting Radiative Neutrino Mass, Neutron-Antineutron Oscillation, Proton Decay, and Leptogenesis through Dark Matter,''
  arXiv:1608.02118 [hep-ph].
  %%CITATION = ARXIV:1608.02118;%%
  %1 citations counted in INSPIRE as of 03 Sep 2016

%\cite{Guo:2016dzl}
\bibitem{Guo:2016dzl} 
  S.~Y.~Guo, Z.~L.~Han and Y.~Liao,
  %``Testing Type II Radiative Seesaw Model: from Dark Matter Detection to LHC Signatures,''
  arXiv:1609.01018 [hep-ph].
  %%CITATION = ARXIV:1609.01018;%% 

 %\cite{Hernandez:2015hrt}
 \bibitem{Hernandez:2015hrt} 
  A.~E.~Carcamo Hernandez,
  %``A novel and economical explanation for SM fermion masses and mixings,''
  arXiv:1512.09092 [hep-ph].
  %%CITATION = ARXIV:1512.09092;%%
  %70 citations counted in INSPIRE as of 07 Sep 2016

  %\cite{Megrelidze:2016fcs}
\bibitem{Megrelidze:2016fcs} 
  L.~Megrelidze and Z.~Tavartkiladze,
  %``Soft See-Saw: Radiative Origin of Neutrino Masses in SUSY Theories,''
  arXiv:1609.07344 [hep-ph].
  %%CITATION = ARXIV:1609.07344;%%

 %\cite{Cheung:2016fjo}
\bibitem{Cheung:2016fjo} 
  K.~Cheung, T.~Nomura and H.~Okada,
  %``A testable radiative neutrino mass model without additional symmetries and Explanation for the $b \to s \ell^+ \ell^-$ anomaly,''
  arXiv:1610.02322 [hep-ph].
  %%CITATION = ARXIV:1610.02322;%%

 %\cite{Seto:2016pks}
\bibitem{Seto:2016pks} 
  O.~Seto and T.~Shimomura,
  %``Atomki anomaly and dark matter in a radiative seesaw model with gauged $B-L$ symmetry,''
  arXiv:1610.08112 [hep-ph].
  %%CITATION = ARXIV:1610.08112;%%
 
 %\cite{Lu:2016dbc}
\bibitem{Lu:2016dbc} 
  W.~B.~Lu and P.~H.~Gu,
  %``Mixed Inert Scalar Triplet Dark Matter, Radiative Neutrino Masses and Leptogenesis,''
  arXiv:1611.02106 [hep-ph].
  %%CITATION = ARXIV:1611.02106;%%
  %2 citations counted in INSPIRE as of 29 Nov 2016
 
 
 
 
 
%%%% Two-loop Begins %%%%%
\bibitem{2-lp-zB} 
%\bibitem{zee-babu}
 A.~Zee,
 %``Quantum Numbers Of Majorana Neutrino Masses,''
 Nucl.\ Phys.\ B {\bf 264}, 99  (1986);
 %%%
 K.~S.~Babu,
 %``MODEL OF 'CALCULABLE' MAJORANA NEUTRINO MASSES,''
 Phys.\ Lett.\ B {\bf 203}, 132  (1988).
 
 

\bibitem{Babu:2002uu} 
  K.~S.~Babu and C.~Macesanu,
  %``Two loop neutrino mass generation and its experimental consequences,''
  Phys.\ Rev.\ D {\bf 67}, 073010 (2003)
  [hep-ph/0212058].
  %%CITATION = HEP-PH/0212058;%%
  %78 citations counted in INSPIRE as of 13 juil. 2015

%\cite{AristizabalSierra:2006gb}
\bibitem{AristizabalSierra:2006gb} 
  D.~Aristizabal Sierra and M.~Hirsch,
  %``Experimental tests for the Babu-Zee two-loop model of Majorana neutrino masses,''
  JHEP {\bf 0612}, 052 (2006)
  [hep-ph/0609307].
  %%CITATION = HEP-PH/0609307;%%
  %60 citations counted in INSPIRE as of 13 Jul 2015

%\cite{Nebot:2007bc}
\bibitem{Nebot:2007bc} 
  M.~Nebot, J.~F.~Oliver, D.~Palao and A.~Santamaria,
  %``Prospects for the Zee-Babu Model at the CERN LHC and low energy experiments,''
  Phys.\ Rev.\ D {\bf 77}, 093013 (2008)
  [arXiv:0711.0483 [hep-ph]].
  %%CITATION = ARXIV:0711.0483;%%
  %61 citations counted in INSPIRE as of 04 juil. 2015

%\cite{Schmidt:2014zoa}
\bibitem{Schmidt:2014zoa} 
  D.~Schmidt, T.~Schwetz and H.~Zhang,
  %``Status of the Zee\UTF{2013}Babu model for neutrino mass and possible tests at a like-sign linear collider,''
  Nucl.\ Phys.\ B {\bf 885}, 524 (2014)
  [arXiv:1402.2251 [hep-ph]].
  %%CITATION = ARXIV:1402.2251;%%
  %11 citations counted in INSPIRE as of 22 Mar 2015

\bibitem{Herrero-Garcia:2014hfa} 
  J.~Herrero-Garcia, M.~Nebot, N.~Rius and A.~Santamaria,
  %``The Zee-Babu model revisited in the light of new data,''
  Nucl.\ Phys.\ B {\bf 885}, 542 (2014)
  [arXiv:1402.4491 [hep-ph]].
  %%CITATION = ARXIV:1402.4491;%%
  %7 citations counted in INSPIRE as of 08 Nov 2014

%\cite{Long:2014fja}
\bibitem{Long:2014fja} 
  H.~N.~Long and V.~V.~Vien,
  %``Neutrino mixing with nonzero $\theta_{13}$ in Zee-Babu model,''
  Int.\ J.\ Mod.\ Phys.\ A {\bf 29}, no. 13, 1450072 (2014)
  [arXiv:1405.1622 [hep-ph]].
  %%CITATION = ARXIV:1405.1622;%%  

%\cite{VanVien:2014apa}
\bibitem{VanVien:2014apa} 
  V.~Van Vien, H.~N.~Long and P.~N.~Thu,
  %``Neutrino mixing and CP violation phases in Zee-Babu model,''
  arXiv:1407.8286 [hep-ph].
  %%CITATION = ARXIV:1407.8286;%%

%\cite{Aoki:2010ib}
\bibitem{Aoki:2010ib} 
  M.~Aoki, S.~Kanemura, T.~Shindou and K.~Yagyu,
  %``An R-parity conserving radiative neutrino mass model without
%right-handed neutrinos,''
  JHEP {\bf 1007}, 084 (2010)
  [Erratum-ibid.\  {\bf 1011}, 049 (2010)]
  [arXiv:1005.5159 [hep-ph]].
  %%CITATION = ARXIV:1005.5159;%%

%\cite{Lindner:2011it}
\bibitem{Lindner:2011it} 
  M.~Lindner, D.~Schmidt and T.~Schwetz,
  %``Dark Matter and Neutrino Masses from Global $U(1)_{B-L}$ Symmetry
%Breaking,''
  Phys.\ Lett.\ B {\bf 705}, 324 (2011)
  [arXiv:1105.4626 [hep-ph]].
  %%CITATION = ARXIV:1105.4626;%%

%\cite{Baek:2012ub}
\bibitem{Baek:2012ub} 
  S.~Baek, P.~Ko, H.~Okada and E.~Senaha,
  %``Can Zee-Babu model implemented with scalar dark matter explain both Fermi/LAT 130 GeV $\gamma$-ray excess and neutrino physics ?,''
  JHEP {\bf 1409}, 153 (2014)
  [arXiv:1209.1685 [hep-ph]].
  %%CITATION = ARXIV:1209.1685;%%
  %22 citations counted in INSPIRE as of 29 Dec 2014

%\cite{Aoki:2013gzs}
\bibitem{Aoki:2013gzs} 
  M.~Aoki, J.~Kubo and H.~Takano,
  %``Two-loop radiative seesaw mechanism with multicomponent dark matter
%explaining the possible \UTF{0160}\UTF{00C3} excess in the Higgs boson decay and
%at the Fermi LAT,''
  Phys.\ Rev.\ D {\bf 87}, no. 11, 116001 (2013)
  [arXiv:1302.3936 [hep-ph]].
  %%CITATION = ARXIV:1302.3936;%%
  %24 citations counted in INSPIRE as of 30 Jul 2014

%\cite{Kajiyama:2013zla}
\bibitem{Kajiyama:2013zla} 
  Y.~Kajiyama, H.~Okada and K.~Yagyu,
  %``Two Loop Radiative Seesaw Model with Inert Triplet Scalar Field,''
  Nucl.\ Phys.\ B {\bf 874}, 198 (2013)
  [arXiv:1303.3463 [hep-ph]].
  %%CITATION = ARXIV:1303.3463;%%
  %10 citations counted in INSPIRE as of 28 Jan 2014

  %\cite{Kajiyama:2013rla}
\bibitem{Kajiyama:2013rla} 
  Y.~Kajiyama, H.~Okada and T.~Toma,
  %``Multicomponent dark matter particles in a two-loop neutrino model,''
  Phys.\ Rev.\ D {\bf 88}, 015029 (2013)
  [arXiv:1303.7356].
  %%CITATION = ARXIV:1303.7356;%%
  %4 citations counted in INSPIRE as of 14 Nov 2013

\bibitem{Baek:2013fsa} 
  S.~Baek, H.~Okada and T.~Toma,
  %``Two loop neutrino model and dark matter particles with global B-L
%symmetry,''
  JCAP {\bf 1406}, 027 (2014)
  [arXiv:1312.3761 [hep-ph]].
  %%CITATION = ARXIV:1312.3761;%%
  %9 citations counted in INSPIRE as of 30 Jul 2014

  %\cite{Okada:2014vla}
\bibitem{Okada:2014vla} 
  H.~Okada,
  %``Two loop Induced Dirac Neutrino Model and Dark Matters with Global $U(1)'$ Symmetry,''
  arXiv:1404.0280 [hep-ph].
  %%CITATION = ARXIV:1404.0280;%%

%\cite{Okada:2014qsa}
\bibitem{Okada:2014qsa} 
  H.~Okada, T.~Toma and K.~Yagyu,
  %``Inert Extension of the Zee-Babu Model,''
  Phys.\ Rev.\ D {\bf 90}, no. 9, 095005 (2014)
  [arXiv:1408.0961 [hep-ph]].
  %%CITATION = ARXIV:1408.0961;%%

%\cite{Okada:2015nga}
\bibitem{Okada:2015nga} 
  H.~Okada,
  %``Two Loop Radiative Seesaw and X-ray line Dark Matter with Global U(1) Symmetry,''
  arXiv:1503.04557 [hep-ph].
  %%CITATION = ARXIV:1503.04557;%%
  %3 citations counted in INSPIRE as of 26 juin 2015

%\cite{Geng:2015sza}
\bibitem{Geng:2015sza} 
  C.~Q.~Geng and L.~H.~Tsai,
  %``Study of Two-Loop Neutrino Mass Generation Models,''
  arXiv:1503.06987 [hep-ph].
  %%CITATION = ARXIV:1503.06987;%%
  %1 citations counted in INSPIRE as of 26 Jun 2015

%\cite{Kashiwase:2015pra}
\bibitem{Kashiwase:2015pra} 
  S.~Kashiwase, H.~Okada, Y.~Orikasa and T.~Toma,
  %``Two Loop Neutrino Model with Dark Matter and Leptogenesis,''
  Int.\ J.\ Mod.\ Phys.\ A {\bf 31}, no. 20n21, 1650121 (2016)
  doi:10.1142/S0217751X16501219
  [arXiv:1505.04665 [hep-ph]].
  %%CITATION = doi:10.1142/S0217751X16501219;%%
  %20 citations counted in INSPIRE as of 29 Nov 2016
  
  
  %\cite{Aoki:2014cja}
\bibitem{Aoki:2014cja} 
  M.~Aoki and T.~Toma,
  %``Impact of semi-annihilation of $\mathbb{Z}_3$ symmetric dark matter
%with radiative neutrino masses,''
  JCAP {\bf 1409}, 016 (2014)
  [arXiv:1405.5870 [hep-ph]].
  %%CITATION = ARXIV:1405.5870;%%
  %4 citations counted in INSPIRE as of 07 Oct 2014

 %\cite{Baek:2014awa}
\bibitem{Baek:2014awa} 
  S.~Baek, H.~Okada and T.~Toma,
  %``Radiative lepton model and dark matter with global $U(1)'$ symmetry,''
  Phys.\ Lett.\ B {\bf 732}, 85 (2014)
  [arXiv:1401.6921 [hep-ph]].
  %%CITATION = ARXIV:1401.6921;%%

%\cite{Okada:2015nca}
\bibitem{Okada:2015nca} 
  H.~Okada and Y.~Orikasa,
  %``Two-loop Neutrino Model with Exotic Leptons,''
  Phys.\ Rev.\ D {\bf 93}, no. 1, 013008 (2016)
  doi:10.1103/PhysRevD.93.013008
  [arXiv:1509.04068 [hep-ph]].
  %%CITATION = doi:10.1103/PhysRevD.93.013008;%%
  %14 citations counted in INSPIRE as of 29 Nov 2016
  
  
  
%\cite{Sierra:2014rxa}
\bibitem{Sierra:2014rxa} 
  D.~Aristizabal Sierra, A.~Degee, L.~Dorame and M.~Hirsch,
  %``Systematic classification of two-loop realizations of the Weinberg operator,''
  JHEP {\bf 1503}, 040 (2015)
  [arXiv:1411.7038 [hep-ph]].
  %%CITATION = ARXIV:1411.7038;%%
  %7 citations counted in INSPIRE as of 22 Mar 2015

  %\cite{Nomura:2016rjf}
\bibitem{Nomura:2016rjf} 
  T.~Nomura and H.~Okada,
  %``Generalized Zee\UTF{2013}Babu model with 750 GeV diphoton resonance,''
  Phys.\ Lett.\ B {\bf 756}, 295 (2016)
%  doi:10.1016/j.physletb.2016.03.034
  [arXiv:1601.07339 [hep-ph]].
  %%CITATION = doi:10.1016/j.physletb.2016.03.034;%%
  %26 citations counted in INSPIRE as of 25 Jun 2016

    %\cite{Nomura:2016run}
\bibitem{Nomura:2016run} 
  T.~Nomura, H.~Okada and Y.~Orikasa,
  %``Radiative Neutrino Mass in Alternative Left-Right Symmetric Model,''
  arXiv:1602.08302 [hep-ph].
  %%CITATION = ARXIV:1602.08302;%%
  %1 citations counted in INSPIRE as of 12 Mar 2016

  %\cite{Bonilla:2016diq}
\bibitem{Bonilla:2016diq} 
  C.~Bonilla, E.~Ma, E.~Peinado and J.~W.~F.~Valle,
  %``Two-loop Dirac neutrino mass and WIMP dark matter,''
  arXiv:1607.03931 [hep-ph].
  %%CITATION = ARXIV:1607.03931;%%

  %\cite{Kohda:2012sr}
\bibitem{Kohda:2012sr} 
  M.~Kohda, H.~Sugiyama and K.~Tsumura,
  %``Lepton number violation at the LHC with leptoquark and diquark,''
  Phys.\ Lett.\ B {\bf 718}, 1436 (2013)
 % doi:10.1016/j.physletb.2012.12.048
  [arXiv:1210.5622 [hep-ph]].
  %%CITATION = doi:10.1016/j.physletb.2012.12.048;%%
  %35 citations counted in INSPIRE as of 07 Aug 2016

  %\cite{Dasgupta:2013cwa}
  \bibitem{Dasgupta:2013cwa} 
  B.~Dasgupta, E.~Ma and K.~Tsumura,
  %``WIMP Dark Matter and Neutrino Mass from Peccei-Quinn Symmetry,''
  Phys.\ Rev.\ D {\bf 89}, 041702 (2014)
  [arXiv:1308.4138 [hep-ph]].
  %%CITATION = ARXIV:1308.4138;%%
  %7 citations counted in INSPIRE as of 30 Jul 2014

%\cite{Nomura:2016ask}
\bibitem{Nomura:2016ask} 
  T.~Nomura and H.~Okada,
  %``An Extended Colored Zee-Babu Model,''
  Phys.\ Rev.\ D {\bf 94}, 075021 (2016)
  doi:10.1103/PhysRevD.94.075021
  [arXiv:1607.04952 [hep-ph]].
  %%CITATION = doi:10.1103/PhysRevD.94.075021;%%
  %6 citations counted in INSPIRE as of 29 Nov 2016
  
  %\cite{Nomura:2016dnf}
\bibitem{Nomura:2016dnf} 
  T.~Nomura, H.~Okada and Y.~Orikasa,
  %``A Radiative Neutrino Model with $SU(2)_L$ Triplet Fields,''
  arXiv:1610.04729 [hep-ph].
  %%CITATION = ARXIV:1610.04729;%%
  %1 citations counted in INSPIRE as of 29 Nov 2016

  
  %\cite{Liu:2016mpf}
\bibitem{Liu:2016mpf} 
  Z.~Liu and P.~H.~Gu,
  %``Extending two Higgs doublet models for two-loop neutrino mass generation and one-loop neutrinoless double beta decay,''
  arXiv:1611.02094 [hep-ph].
  %%CITATION = ARXIV:1611.02094;%%
  
  
  
  

  
  
  %%%% Three-loop Begins %%%%%
%\cite{Krauss:2002px}
\bibitem{Krauss:2002px}
  L.~M.~Krauss, S.~Nasri and M.~Trodden,
  %``A Model for neutrino masses and dark matter,''
  Phys.\ Rev.\  D {\bf 67}, 085002 (2003)
  [arXiv:hep-ph/0210389].
  %%CITATION = PHRVA,D67,085002;%%

%\cite{Aoki:2008av}
\bibitem{Aoki:2008av}
  M.~Aoki, S.~Kanemura and O.~Seto,
  %``Neutrino mass, Dark Matter and Baryon Asymmetry via TeV-Scale
	%Physics
  %without Fine-Tuning,''
  Phys.\ Rev.\ Lett.\  {\bf 102}, 051805 (2009)
  [arXiv:0807.0361].
  %%CITATION = PRLTA,102,051805;%%

  %\cite{Gustafsson}
\bibitem{Gustafsson:2012vj} 
M.~Gustafsson, J.~M.~No and M.~A.~Rivera,
%``The Cocktail Model: Neutrino Masses and Mixings with Dark Matter''
  Phys.\ Rev.\ Lett.\  {\bf 110}, 211802 (2013)
arXiv:1212.4806 [hep-ph].
  %3 citations counted in INSPIRE as of 19 Sep 2013

%\cite{Ahriche:2014xra}
\bibitem{Ahriche:2014xra} 
  A.~Ahriche, S.~Nasri and R.~Soualah,
  %``Radiative Neutrino Mass Model at the $e^{-}e^{+}$ Linear
%Collider,''
  Phys.\ Rev.\ D {\bf 89}, 095010 (2014)
  [arXiv:1403.5694 [hep-ph]].
  %%CITATION = ARXIV:1403.5694;%%
  %7 citations counted in INSPIRE as of 30 Jul 2014

%\cite{Ahriche:2014cda}
\bibitem{Ahriche:2014cda} 
  A.~Ahriche, C.~S.~Chen, K.~L.~McDonald and S.~Nasri,
  %``Three-loop model of neutrino mass with dark matter,''
  Phys.\ Rev.\ D {\bf 90}, no. 1, 015024 (2014)
  [arXiv:1404.2696 [hep-ph]].
  %%CITATION = ARXIV:1404.2696;%%
  %18 citations counted in INSPIRE as of 22 mar 2015

%\cite{Ahriche:2014oda}
\bibitem{Ahriche:2014oda} 
  A.~Ahriche, K.~L.~McDonald and S.~Nasri,
  %``A Model of Radiative Neutrino Mass: with or without Dark Matter,''
  JHEP {\bf 1410}, 167 (2014)
  [arXiv:1404.5917 [hep-ph]].
  %%CITATION = ARXIV:1404.5917;%%
  %15 citations counted in INSPIRE as of 22 mar 2015

%\cite{Okada:2014oda}
\bibitem{Okada:2014oda} 
  H.~Okada and Y.~Orikasa,
  %``X-ray line in Radiative Neutrino Model with Global $U(1)$ Symmetry,''
  Phys.\ Rev.\ D {\bf 90}, no. 7, 075023 (2014)
  [arXiv:1407.2543 [hep-ph]].
  %%CITATION = ARXIV:1407.2543;%%
  %9 citations counted in INSPIRE as of 10 Dec 2014

%\cite{Hatanaka:2014tba}
\bibitem{Hatanaka:2014tba} 
  H.~Hatanaka, K.~Nishiwaki, H.~Okada and Y.~Orikasa,
  %``A Three-Loop Neutrino Model with Global $U(1)$ Symmetry,''
  Nucl.\ Phys.\ B {\bf 894}, 268 (2015)
  [arXiv:1412.8664 [hep-ph]].
  %%CITATION = ARXIV:1412.8664;%%
  %7 citations counted in INSPIRE as of 21 Apr 2015

%\cite{Jin:2015cla}
\bibitem{Jin:2015cla} 
  L.~G.~Jin, R.~Tang and F.~Zhang,
  %``A three-loop radiative neutrino mass model with dark matter,''
  Phys.\ Lett.\ B {\bf 741}, 163 (2015)
  [arXiv:1501.02020 [hep-ph]].
  %%CITATION = ARXIV:1501.02020;%%

%\cite{Culjak:2015qja}
\bibitem{Culjak:2015qja} 
  P.~Culjak, K.~Kumericki and I.~Picek,
  %``Scotogenic R$\nu$MDM at three-loop level,''
  Phys.\ Lett.\ B {\bf 744}, 237 (2015)
  [arXiv:1502.07887 [hep-ph]].
  %%CITATION = ARXIV:1502.07887;%%
  %2 citations counted in INSPIRE as of 26 juin 2015

%\cite{Okada:2015bxa}
\bibitem{Okada:2015bxa} 
H.~Okada, N.~Okada and Y.~Orikasa,
%``Radiative seesaw mechanism in a minimal 3-3-1 model,’'
Phys.\ Rev.\ D {\bf 93}, no. 7, 073006 (2016)
%doi:10.1103/PhysRevD.93.073006
[arXiv:1504.01204 [hep-ph]].

%\cite{Geng:2015coa}
\bibitem{Geng:2015coa} 
  C.~Q.~Geng, D.~Huang and L.~H.~Tsai,
  %``Comment on "A three-loop radiative neutrino mass model with dark matter" [Phys. Lett. B 741 (2015) 163],''
  Phys.\ Lett.\ B {\bf 745}, 56 (2015)
  [arXiv:1504.05468 [hep-ph]].
  %%CITATION = ARXIV:1504.05468;%%
  %1 citations counted in INSPIRE as of 26 Jun 2015

%\cite{Ahriche:2015wha}
\bibitem{Ahriche:2015wha} 
  A.~Ahriche, K.~L.~McDonald, S.~Nasri and T.~Toma,
  %``A Model of Neutrino Mass and Dark Matter with an Accidental Symmetry,''
  Phys.\ Lett.\ B {\bf 746}, 430 (2015)
  [arXiv:1504.05755 [hep-ph]].
  %%CITATION = ARXIV:1504.05755;%%
  %4 citations counted in INSPIRE as of 26 juin 2015

%\cite{Nishiwaki:2015iqa}
\bibitem{Nishiwaki:2015iqa} 
  K.~Nishiwaki, H.~Okada and Y.~Orikasa,
  %``Three loop neutrino model with isolated $k^{\pm\pm}$,''
  Phys.\ Rev.\ D {\bf 92}, no. 9, 093013 (2015)
  doi:10.1103/PhysRevD.92.093013
  [arXiv:1507.02412 [hep-ph]].
  %%CITATION = doi:10.1103/PhysRevD.92.093013;%%
  %24 citations counted in INSPIRE as of 29 Nov 2016
  
  
%\cite{Okada:2015hia}
\bibitem{Okada:2015hia} 
  H.~Okada and K.~Yagyu,
  %``Three-loop neutrino mass model with doubly charged particles from isodoublets,''
  Phys.\ Rev.\ D {\bf 93}, no. 1, 013004 (2016)
  doi:10.1103/PhysRevD.93.013004
  [arXiv:1508.01046 [hep-ph]].
  %%CITATION = doi:10.1103/PhysRevD.93.013004;%%
  %17 citations counted in INSPIRE as of 29 Nov 2016
  
  
%\cite{Ahriche:2015loa}
\bibitem{Ahriche:2015loa} 
  A.~Ahriche, K.~L.~McDonald and S.~Nasri,
  %``A Radiative Model for the Weak Scale and Neutrino Mass via Dark Matter,''
  arXiv:1508.02607 [hep-ph].
  %%CITATION = ARXIV:1508.02607;%%
  
%\cite{Kajiyama:2013lja}
\bibitem{Kajiyama:2013lja} 
  Y.~Kajiyama, H.~Okada and K.~Yagyu,
  %``$T_7$ Flavor Model in Three Loop Seesaw and Higgs Phenomenology,''
  JHEP {\bf 10}, 196 (2013)
  arXiv:1307.0480 [hep-ph].
  %%CITATION = ARXIV:1307.0480;%%

  %\cite{King:2014uha}
\bibitem{King:2014uha} 
  S.~F.~King, A.~Merle and L.~Panizzi,
  %``Effective theory of a doubly charged singlet scalar: complementarity of neutrino physics and the LHC,''
  arXiv:1406.4137 [hep-ph].
  %%CITATION = ARXIV:1406.4137;%%
  %1 citations counted in INSPIRE as of 31 Aug 2014

%\cite{Kanemura:2015bli}
\bibitem{Kanemura:2015bli} 
  S.~Kanemura, K.~Nishiwaki, H.~Okada, Y.~Orikasa, S.~C.~Park and R.~Watanabe,
  %``LHC 750 GeV Diphoton excess in a radiative seesaw model,''
  arXiv:1512.09048 [hep-ph].
  %%CITATION = ARXIV:1512.09048;%%
  %68 citations counted in INSPIRE as of 25 Jun 2016

  %\cite{Okada:2016rav}
\bibitem{Okada:2016rav} 
  H.~Okada and K.~Yagyu,
  %``Renormalizable model for neutrino mass, dark matter, muon $g-2$ and 750 GeV diphoton excess,''
  Phys.\ Lett.\ B {\bf 756}, 337 (2016)
%  doi:10.1016/j.physletb.2016.03.040
  [arXiv:1601.05038 [hep-ph]].
  %%CITATION = doi:10.1016/j.physletb.2016.03.040;%%
  %29 citations counted in INSPIRE as of 25 Jun 2016

%\cite{Ko:2016sxg}
 \bibitem{Ko:2016sxg} 
  P.~Ko, T.~Nomura, H.~Okada and Y.~Orikasa,
  %``Confronting a New Three-loop Seesaw Model with the 750 GeV Diphoton Excess,''
  arXiv:1602.07214 [hep-ph].
  %%CITATION = ARXIV:1602.07214;%%
  %14 citations counted in INSPIRE as of 25 Jun 2016

%\cite{Nomura:2016vxr}
\bibitem{Nomura:2016vxr} 
  T.~Nomura, H.~Okada and Y.~Orikasa,
  %``Radiative Seesaw Model with Degenerate Majorana Dark Matter,''
  Phys.\ Rev.\ D {\bf 93}, no. 11, 113008 (2016)
  doi:10.1103/PhysRevD.93.113008
  [arXiv:1603.04631 [hep-ph]].
  %%CITATION = doi:10.1103/PhysRevD.93.113008;%%
  %11 citations counted in INSPIRE as of 29 Nov 2016
  
  
  %\cite{Thuc:2016qva}
\bibitem{Thuc:2016qva} 
  T.~T.~Thuc, L.~T.~Hue, H.~N.~Long and T.~P.~Nguyen,
  %``Lepton flavor violating decay of SM-like Higgs in a radiative neutrino mass model,''
  arXiv:1604.03285 [hep-ph].
  %%CITATION = ARXIV:1604.03285;%%
  %1 citations counted in INSPIRE as of 08 May 2016

 %\cite{Cherigui:2016tbm}
\bibitem{Cherigui:2016tbm} 
  D.~Cherigui, C.~Guella, A.~Ahriche and S.~Nasri,
  %``Probing Radiative Neutrino Mass Models Using Trilepton Channel at the LHC,''
  arXiv:1605.03640 [hep-ph].
  %%CITATION = ARXIV:1605.03640;%%

%\cite{Nomura:2016ezz}
\bibitem{Nomura:2016ezz} 
  T.~Nomura, H.~Okada and N.~Okada,
  %``A Colored KNT Neutrino Model,''
  Phys.\ Lett.\ B {\bf 762}, 409 (2016)
  doi:10.1016/j.physletb.2016.09.038
  [arXiv:1608.02694 [hep-ph]].
  %%CITATION = doi:10.1016/j.physletb.2016.09.038;%%
  %6 citations counted in INSPIRE as of 29 Nov 2016  
  
%\cite{Cheung:2016ypw}
\bibitem{Cheung:2016ypw} 
  K.~Cheung, H.~Ishida and H.~Okada,
  %``Accommodation of the Dirac Phase in the Krauss-Nasri-Trodden Model,''
  arXiv:1609.06231 [hep-ph].
  %%CITATION = ARXIV:1609.06231;%%
  
  %\cite{Cheung:2016frv}
\bibitem{Cheung:2016frv} 
  K.~Cheung, T.~Nomura and H.~Okada,
  %``A Three-loop Neutrino Mass Model with a Colored Triplet Scalar,''
  arXiv:1610.04986 [hep-ph].
  %%CITATION = ARXIV:1610.04986;%%

%\cite{Gu:2016xno}
\bibitem{Gu:2016xno} 
  P.~H.~Gu,
  %``High-scale leptogenesis with three-loop neutrino mass generation and dark matter,''
  arXiv:1611.03256 [hep-ph].
  %%CITATION = ARXIV:1611.03256;%%
  
  
  
  
  
  
  


%%%% Four-loop Begins %%%%%
  %\cite{Nomura:2016fzs}
\bibitem{Nomura:2016fzs} 
  T.~Nomura and H.~Okada,
  %``Four-loop Neutrino Model Inspired by Diphoton Excess at 750 GeV,''
  Phys.\ Lett.\ B {\bf 755}, 306 (2016)
%  doi:10.1016/j.physletb.2016.02.022
  [arXiv:1601.00386 [hep-ph]].
   %%CITATION = doi:10.1016/j.physletb.2016.02.022;%%
  %49 citations counted in INSPIRE as of 08 May 2016
  
  %\cite{Nomura:2016seu}
\bibitem{Nomura:2016seu} 
  T.~Nomura and H.~Okada,
  %``Four-loop Radiative Seesaw Model with 750 GeV Diphoton Resonance,''
  arXiv:1601.04516 [hep-ph].
  %%CITATION = ARXIV:1601.04516;%%
  %37 citations counted in INSPIRE as of 08 May 2016







%\cite{Forero:2014bxa}
\bibitem{Forero:2014bxa} 
  D.~V.~Forero, M.~Tortola and J.~W.~F.~Valle,
  %``Neutrino oscillations refitted,''
  Phys.\ Rev.\ D {\bf 90}, no. 9, 093006 (2014)
%  doi:10.1103/PhysRevD.90.093006
  [arXiv:1405.7540 [hep-ph]].
  %%CITATION = doi:10.1103/PhysRevD.90.093006;%%
  %346 citations counted in INSPIRE as of 01 Sep 2016
  

  
   %\cite{TheMEG:2016wtm}
\bibitem{TheMEG:2016wtm} 
  A.~M.~Baldini {\it et al.} [MEG Collaboration],
  %``Search for the Lepton Flavour Violating Decay $\mu^{+} \to e^+ \gamma$ with the Full Dataset of the MEG Experiment,''
  arXiv:1605.05081 [hep-ex].
  %%CITATION = ARXIV:1605.05081;%%
  %4 citations counted in INSPIRE as of 25 Jun 2016
  
  
  
%\cite{Adam:2013mnn}
\bibitem{Adam:2013mnn} 
  J.~Adam {\it et al.} [MEG Collaboration],
  %``New constraint on the existence of the $\mu^+ \to e^+\gamma$ decay,''
  Phys.\ Rev.\ Lett.\  {\bf 110}, 201801 (2013)
  [arXiv:1303.0754 [hep-ex]].
  %%CITATION = ARXIV:1303.0754;%%
  %297 citations counted in INSPIRE as of 08 sept. 2015

  
  %\cite{Akerib:2016vxi}
\bibitem{Akerib:2016vxi} 
  D.~S.~Akerib {\it et al.},
  %``Results from a search for dark matter in LUX with 332 live days of exposure,''
  arXiv:1608.07648 [astro-ph.CO].
  %%CITATION = ARXIV:1608.07648;%%
  %1 citations counted in INSPIRE as of 03 Sep 2016
  
      %\cite{Arhrib:2013ela}
\bibitem{Arhrib:2013ela} 
  A.~Arhrib, Y.~L.~S.~Tsai, Q.~Yuan and T.~C.~Yuan,
  %``An Updated Analysis of Inert Higgs Doublet Model in light of the Recent Results from LUX, PLANCK, AMS-02 and LHC,''
  JCAP {\bf 1406}, 030 (2014)
%  doi:10.1088/1475-7516/2014/06/030
  [arXiv:1310.0358 [hep-ph]].
  %%CITATION = doi:10.1088/1475-7516/2014/06/030;%%
  %66 citations counted in INSPIRE as of 05 Sep 2016
  
  %\cite{Griest:1990kh}
\bibitem{Griest:1990kh} 
  K.~Griest and D.~Seckel,
  %``Three exceptions in the calculation of relic abundances,''
  Phys.\ Rev.\ D {\bf 43}, 3191 (1991).
%  doi:10.1103/PhysRevD.43.3191
  %%CITATION = doi:10.1103/PhysRevD.43.3191;%%
  %796 citations counted in INSPIRE as of 03 Sep 2016
  
  %\cite{Edsjo:1997bg}
\bibitem{Edsjo:1997bg} 
  J.~Edsjo and P.~Gondolo,
  %``Neutralino relic density including coannihilations,''
  Phys.\ Rev.\ D {\bf 56}, 1879 (1997)
%  doi:10.1103/PhysRevD.56.1879
  [hep-ph/9704361].
  %%CITATION = doi:10.1103/PhysRevD.56.1879;%%
  %446 citations counted in INSPIRE as of 03 Sep 2016
  
  %\cite{Ade:2013zuv}
\bibitem{Ade:2013zuv} 
  P.~A.~R.~Ade {\it et al.} [Planck Collaboration],
  %``Planck 2013 results. XVI. Cosmological parameters,''
  Astron.\ Astrophys.\  {\bf 571}, A16 (2014)
%  doi:10.1051/0004-6361/201321591
  [arXiv:1303.5076 [astro-ph.CO]].
  %%CITATION = doi:10.1051/0004-6361/201321591;%%
  %4826 citations counted in INSPIRE as of 03 Sep 2016
  
  %\cite{Bennett:2006fi}
\bibitem{Bennett:2006fi} 
  G.~W.~Bennett {\it et al.} [Muon g-2 Collaboration],
  %``Final Report of the Muon E821 Anomalous Magnetic Moment Measurement at BNL,''
  Phys.\ Rev.\ D {\bf 73}, 072003 (2006)
 % doi:10.1103/PhysRevD.73.072003
  [hep-ex/0602035].
  %%CITATION = doi:10.1103/PhysRevD.73.072003;%%
  %1239 citations counted in INSPIRE as of 03 Sep 2016
  

  
\end{thebibliography}
\end{document}